\documentclass[aps, prd, floats, floatfix, twocolumn,superscriptaddress, nofootinbib]{revtex4-1}

\usepackage{graphicx}
\usepackage{hyperref}
\usepackage{amsmath,amssymb}
\usepackage{amsfonts}
\usepackage{xspace} 
\usepackage[usenames]{color}
\usepackage{dcolumn} 
\usepackage{bm} 
\usepackage{epsfig}
\usepackage{aas_macros}

\definecolor {darkgreen}{rgb}{0.2,0.7,0.2}

\newcommand{\be}{\begin{equation}}
\newcommand{\ba}{\begin{eqnarray}}
\newcommand{\ee}{\end{equation}}
\newcommand{\ea}{\end{eqnarray}}

\newcommand{\SMBH}{\bullet}
\newcommand{\BH}{\mathrm{\bullet}}
\newcommand{\Gpc}{\,\mathrm{Gpc}}
\newcommand{\Mpc}{\,\mathrm{Mpc}}

\newcommand{\Msun}{\,{\rm M_\odot}}
\newcommand{\yr}{\,{\rm yr}}

\newcommand{\Hz}{\,{\rm Hz}}

\newcommand{\D}{\mathrm{d}}

\newcommand{\dL}{d_{\rm L}}

\begin{document}

\title{Repeated Bursts from Relativistic Scattering of Compact Objects in Galactic Nuclei}

\author{Bence Kocsis}
\affiliation{Harvard-Smithsonian Center for Astrophysics, 60 Garden St., Cambridge, MA 02138, USA.}
\affiliation{Einstein Fellow}

\author{Janna Levin}
\affiliation{Department of Physics and Astronomy, Barnard College of Columbia University, 3009 Broadway, New York, NY 10027}
\affiliation{Institute for Strings, Cosmology and Astroparticle Physics, Columbia University, New York, NY 10027}

\begin{abstract}
Galactic nuclei are densely populated by stellar mass compact objects such as black holes and neutron stars. Bound, highly eccentric binaries form as a result of gravitational wave (GW) losses during close flybys between these objects. We study the evolution of these systems using 2.5 and 3.5 order post-Newtonian equations of motion. The GW signal consists of many thousand repeated bursts (RB) for minutes to days (depending on the impact parameter and masses), followed by a powerful GW chirp and an eccentric merger. We show that a significant signal to noise ratio (SNR) accumulates already in the RB phase, corresponding to a detection limit around 200--300 Mpc and 300--600 Mpc for Advanced LIGO for an average orientation BH/NS or BH/BH binary, respectively. The theoretical errors introduced by the inaccuracy of the PN templates are typically much less severe for the RB phase than in the following eccentric merger. The GW signal in the RB phase is broadband; we show that encounters involving intermediate mass black holes are detectable in multiple frequency bands coincidentally using LIGO and LISA.
\end{abstract}
\date{\today \hspace{0.2truecm}}
\maketitle

\section{Introduction}

Close approaches between initially unbound compact objects (COs) can form bound binary systems if the
gravitational wave (GW) emission, tidal dissipation, or interaction
with other objects taps enough of the initial kinetic energy
\cite{2006ApJ...648..411K,2009MNRAS.395.2127O,2010ApJ...720..953L}.
In particular, GW captures of black holes (BHs) and neutron stars (NSs) occur many times per Hubble time
in dense stellar environments like galactic nuclei or globular
clusters. These pairs provide sources for
direct GW detection with Advaced LIGO or Virgo
\cite{2006ApJ...648..411K,2009MNRAS.395.2127O}, and can lead to short-hard gamma ray bursts (SGRBs)
\cite{2005MNRAS.356...54D,2010ApJ...720..953L,2011arXiv1105.3175S,2012PhRvL.108a1102T}.

The GW signals of these eccentric sources are very different from standard quasi-circular inspirals.
According to the leading order results \cite{1963PhRv..131..435P}, for a given
semimajor axis they are more luminous and are described by broadband spectra,
which makes them detectable to larger distances and in a broader mass-range.
Kocsis, G\'asp\'ar, \& M\'{a}rka (hereafter KGM) \cite{2006ApJ...648..411K} investigated
the detectability of the GW burst emitted
during a single passage, and found the SNR to be substantial only for encounters with a very small initial pericenter
distance $r_{p0} \lesssim 6 M$ (where $M$ is the total mass in units $G=c=1$), which occurs relatively rarely.
O'Leary, Kocsis, \& Loeb (hereafter OKL)
\cite{2009MNRAS.395.2127O} included the much stronger GW signal from subsequent passages from bound systems following the
first encounter, leading to an eccentric inspiral. Remarkably, the expected detection rates of these sources for Advanced LIGO is
comparable to other types of waveforms, between
$1$--$10^3\yr^{-1}$. The large uncertainty is mostly due to the unknown number and mass distribution of BHs in galactic nuclei
{ (see Appendix~\ref{app:rates} for further discussion)}.

As the binary evolves from the initial very eccentric phase towards the less eccentric phase,
the GW signal initially consists of well-separated repeated bursts (RBs) for minutes to days,
and later transitions to a continuous inspiral waveform, a short but powerful chirp (OKL). The signal evolves from the RB to the chirp phase
within the frequency band of Advanced-LIGO type instruments, making these GW signals particularly rich in features
and very unique among other sources. These GW sources, if involving NSs, have electromagnetic counterparts, making
them interesting candidates for multimessenger astronomy \cite{2011CQGra..28k4013M,McWilliams:2011zi}.

Existing techniques are not well suited to dig these GW signals out of the noise,
in either the RB or the final chirp phase. In the RB phase,
individual GW bursts are relatively weak compared to the instrumental noise, making burst search algorithms insensitive
to these sources. Nevertheless, since the time evolution of successive bursts can be predicted theoretically,
and there are hundreds to thousands of well-separated bursts in the RB phase, it is in priniciple possible to
optimize detection algorithms to coherently detect the full sequence of bursts. Regarding the final chirp,
existing matched filtering searches with circular inspiral templates are also expected to be ineffective, as here the eccentricity is still considerable
\cite{1999PhRvD..60l4008M,2009CQGra..26d5013C,2010PhRvD..81b4007B}.
Post-Newtonian (PN) or effective one body (EOB) body waveforms have not been developed to sufficient accuracy
for eccentric orbits of comparable-mass binaries with a small pericenter distance
\cite{2010PhRvD..82b4033H,2010PhRvD..81b4017D}.
Direct numerical experiments are
restricted to non-extreme eccentricities ($e<0.7$) and only a limited number of configurations have been tested
\cite{2007CQGra..24...83P,2008PhRvD..77h1502H,2008PhRvD..78f4069S,2009PhRvL.103m1101H}.
Without sufficiently accurate theoretical templates, matched filtering detection techniques will be prone to large
theoretical errors \cite{2007PhRvD..76j4018C}. These issues might be expected to be less severe in the RB phase,
where the binary separation is relatively large, as long as the source is not in the zoom-whirl regime
 \cite{2008PhRvD..77j3005L,2009PhRvD..79d3016L,2009PhRvD..79d3017G,1994PhRvD..50.3816C,2002PhRvD..66d4002G,2010PhRvD..81h4021B}.

In this paper, we focus on the detectability of the GWs emitted in the RB phase.
We examine the relativistic corrections to the evolution of the GW-capture binaries.
We numerically integrate the 2.5PN and 3.5PN equations of motion of the binary,
including the radiation-reaction force.
It is important to note that we do not include trajectories that
technically whirl (execute a full $2\pi$ or more around periastron)
since these orbits are by necessity in a regime where the PN equations
of motion are unreliable. Numerical relativity is needed to examine
additional boosts to SNR due to whirls.
We calculate the GWs emitted  and evaluate the
numerical Fast Fourier Transform (FFT),
to compare with the detection threshold of GW instruments. We study how the SNR accumulates
in time during the evolution, and examine whether these broadband waveforms can be detected coincidentally
in separate frequency channels with different GW instruments. (We find
that they can.)
{ We provide a brief estimate of the event rates in the Appendix.}

We use units $G=c=1$.

\section{Evolution of orbits}

We integrate the instananeous 2.5PN and 3.5PN equations of motion of Will and collaborators, including spin corrections
and dissipation due to gravitational radiation emission
\cite{2002PhRvD..65j4008P,2004PhRvD..69j4021M,2005PhRvD..71l9901M,2005PhRvD..71h4027W,2007PhRvD..75f4017W}.
Simultaneously to numerical integration of the trajectories, we calculate the two
GW polarizations. Note that this is different from the approach used for quasicircular orbits, where the orbit-averaged
fluxes are calculated to a much higher order: 3PN order beyond the 2.5PN leading order flux
\cite{2007PhRvD..76l4038B,2007PhRvL..99r1101B,2007PhRvD..75l4018B,2010PhRvD..82b4033H}.
Direct integration of the equations of motion allows greater
flexibility when working with very high eccentricities, where averaging the GW flux over a Newtonian approximation to the
orbital geometry would be very inaccurate.
\footnote{{ Our approach is also different from that used in \citet{2004PhRvD..70f4028D} or \citet{2009PhRvD..80l4018A},
which give the phasing of binaries to 3.5PN order by averaging the radiation reaction over an
orbital period and using this in a calculation of angular momentum
flux. Orbit-averaged fluxes are computed to higher orders
than explicit equations of motion. As a result, variations in the
signs of radiation-reaction terms at different orders
can be washed out with orbital averaging. However, the
limits of validity of the PN expansion are still pressed at close radial
separations, regardless of the approach.}}

We use the 2.5PN and 3.5PN approximations to investigate the relativistic corrections to the OKL study,
and to assess the calculational uncertainties.
While the 3.5PN approximation is more accurate than the 2.5PN
approximation at large separations,
the PN calculation breaks down interior to $r_{p}\lesssim 10 M$ in
this calculation scheme,
as in this case the 3.5PN correction dominates over the 2.5PN terms and leads to an artificial increase of
the eccentricity \cite{2011CQGra..28q5001L}.
The PN approximation is not well behaved in this important regime.
However, the binding energy decreases monotonically until merger for the 2.5PN calculation.
For the 3.5PN runs, we terminate the simulation where the magnitude of the 3.5PN perturbation terms
first dominate over the lower-order terms. This is usually in the final chirp phase.
Predictions for the RB phase are typically not affected by this truncation, as long as the binary
is not in the zoom-whirl regime.

\subsection{Binary formation}
\begin{figure*}
\centering{
\mbox{\includegraphics{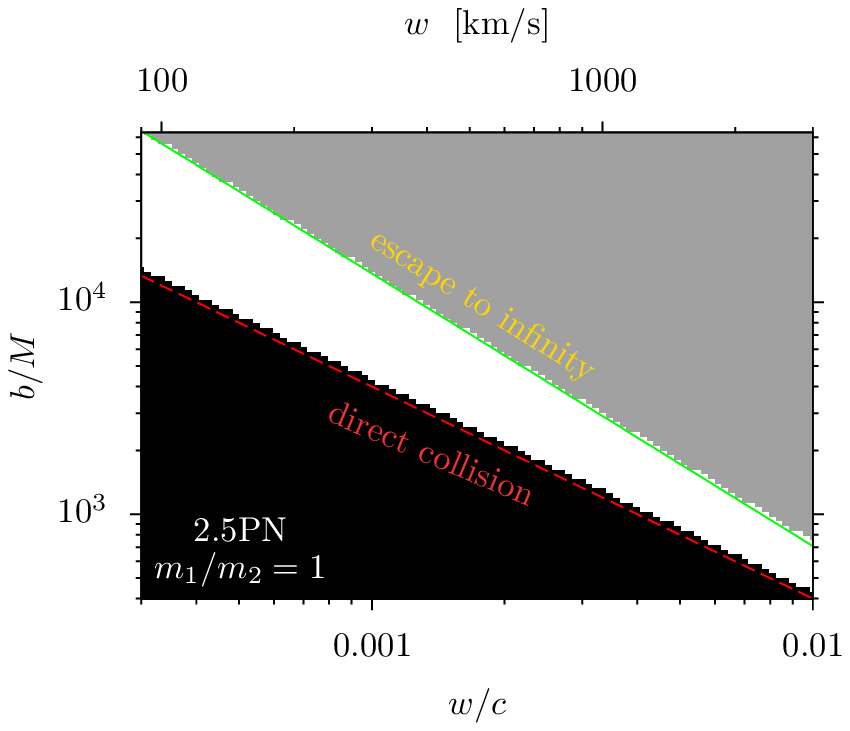}}
\mbox{\includegraphics{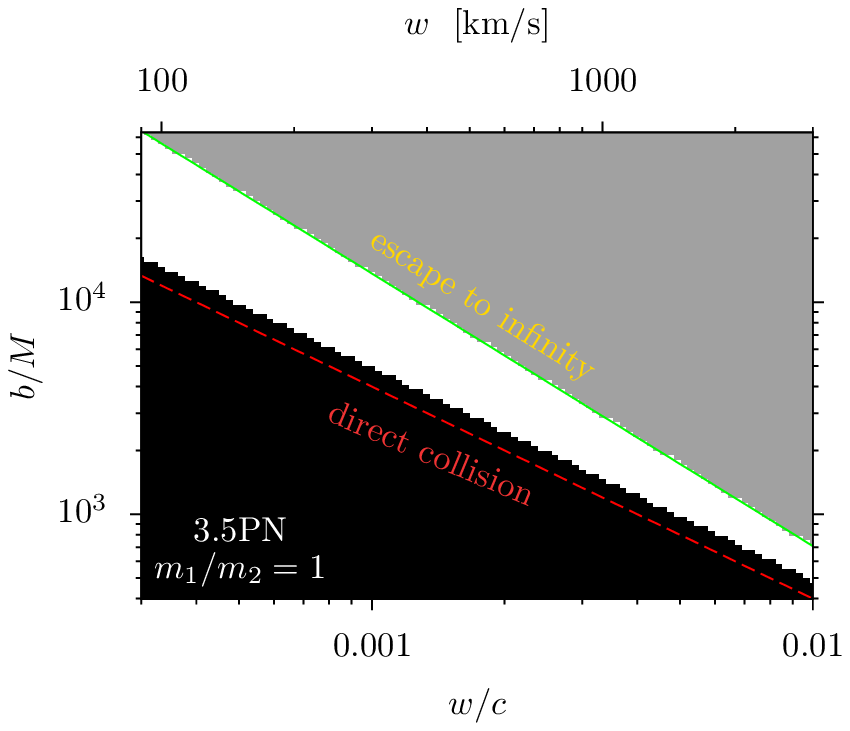}}\\\vspace{3pt}
\mbox{\includegraphics{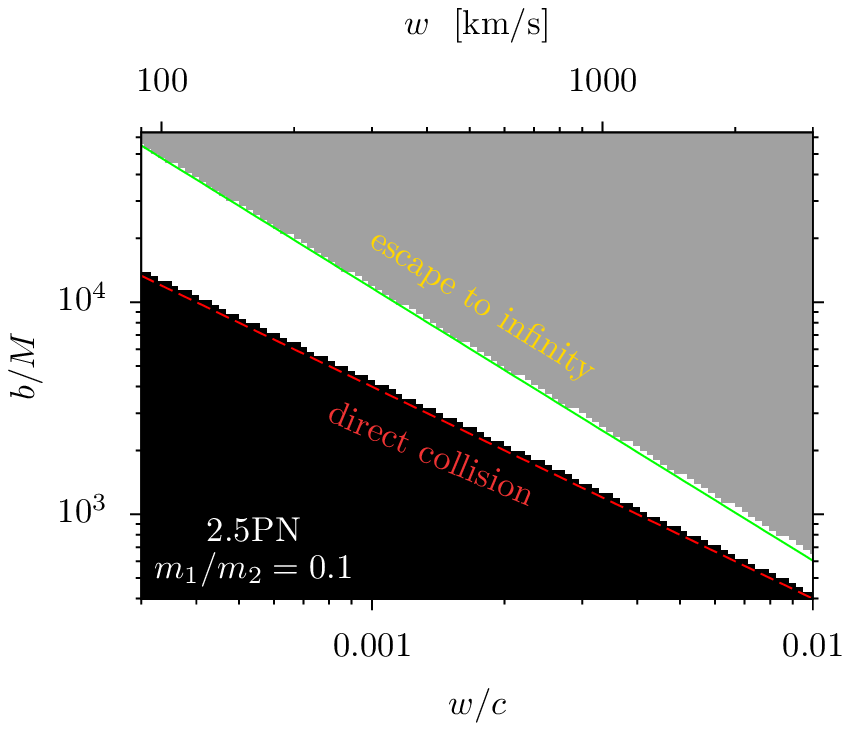}}
\mbox{\includegraphics{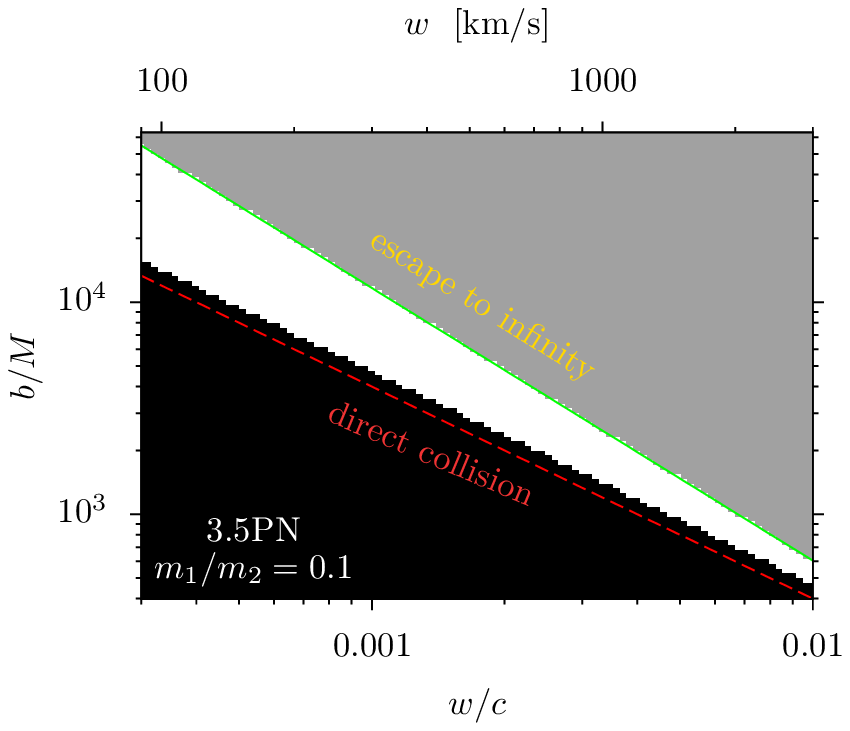}}
}
\caption{Possible outcomes of the encounters depending on the initial velocity and impact parameter: escape (gray),
capture into bound elliptic orbits (white), or direct collision (black). { Lines show the analytical estimates based on Eq.~(\ref{eq:bN}).
Top and bottom panels are for mass ratio 1 and 0.1,} left and right panels correspond to 2.5PN and 3.5PN calculations, respectively.
The 3.5PN calculations were terminated at radii where the derived perturbations are unphysical in the black region.
\label{fig:basin}}
\end{figure*}
There are three possible outcomes after the first close encounter between two compact objects,
depending on the initial conditions:
unbound quasi-hyperbolic trajectory, capture into bound quasi-eccentric orbit, or direct collision. Here we examine
relativistic corrections to the capture cross section.

The event rates of these waveforms are sensitive to the critical impact parameter for capture into a bound orbit
$b_{\max}=b_{\max}(m_1,m_2,w)$. Here $m_1$ and $m_2$ are the component masses, and $w$ is the initial relative velocity at infinity.
To leading order, a bound (non-plunging) system forms if
\begin{equation}
\frac{4 M}{w}\lesssim b\lesssim \left (\frac{340\pi}{3}\right
)^{1/7}M\frac{\eta^{1/7}}{w^{9/7}}\,.
\label{eq:bN}
\end{equation}
where $M=m_1+m_2$.
Here the upper bound assumes quadrupolar radiation emitted on a hyperbolic trajectory \cite{1964PhRv..136.1224P,1977ApJ...216..610T},
and the lower bound is valid in the test-particle limit around a Schwarzschild BH (see KGM and OKL).

Figure~\ref{fig:basin} shows the boundaries for 2.5PN and 3.5PN simulations for equal masses { and for mass ratios $m_1/m_2=0.1$} and no spins.
The green line shows that Eq.~(\ref{eq:bN}) used by OKL for binary formation
is in excellent agreement with our numerical post-Newtonian calculation { for these mass ratios.  This is also expected from analytical orbital-averaged
PN estimates for the typically nonrelativistic initial velocities in galactic nuclei ($w\ll 0.01$).}\footnote{{
The post-Newtonian correction to the RHS of Eq.~(\ref{eq:bN}) was calculated by \citet{1992MNRAS.254..146J}.
Expanding their Eq.~(48) in a power series in $w$ to first to leading order gives
\begin{equation}
b_{3.5\rm PN}^{\rm max} \approx b_{2.5\rm PN}^{\rm max} \left[ 1 + \frac{ (5763-3220\eta)}{3400} \left(\frac{3}{340\pi \eta}\right)^{2/7}w^{4/7} \right]
\end{equation}
which yields deviations from the leading order term by less than $3\%$ for equal masses $\eta=1/4$ and $w\leq 0.01$.
The correction is larger only for very unequal mass ratios, but such sources are not expected to exist for terrestrial GW instruments,
based on the frequency limit of the instrument (implying $M \lesssim 100 \Msun$) and the minimum mass of BHs and NSs ($M\gtrsim 1\Msun$).
}
}
We are unable to resolve orbits that cross interior to $10 M$, even if they are not direct captures, as in this regime
higher order PN effects may be more significant \cite{2011CQGra..28q5001L}.
In the following we focus on orbits that are captured in bound eccentric orbits.

{
Combining the above estimates for the impact parameter with the expected number density of objects in the galactic nucleus
and their velocity distribution can be used to make estimates on the likelihood of such encounters.
This excersize, summarized in Appendix~\ref{app:rates}, yields that out of $10^4$ compact objects in a single galactic nucleus,
only a few binaries form in a billion years. The corresponding instantaneous fraction of objects in the binary forming region
shown in Fig.~\ref{fig:basin} is extremely small on average. However, if the detectable distance of these sources is sufficiently large,
the total rate from all observable galaxies may be quite high.
}

We conclude that relativistic corrections do not modify the capture cross section over the $10\%$ uncertainty in the simple OKL estimate
associated to the relative velocity distribution. { These relativistic corrections are negligible compared to the
theoretical uncertainties in the event rates as discussed in Appendix~\ref{app:rates}. The detection rates however
may be affected by relativistic corrections through variations in the detectable distance of the source, which we investigate below.
}

\subsection{Orbital evolution}
\begin{figure*}
\centering{
\mbox{\includegraphics[width=8.5cm]{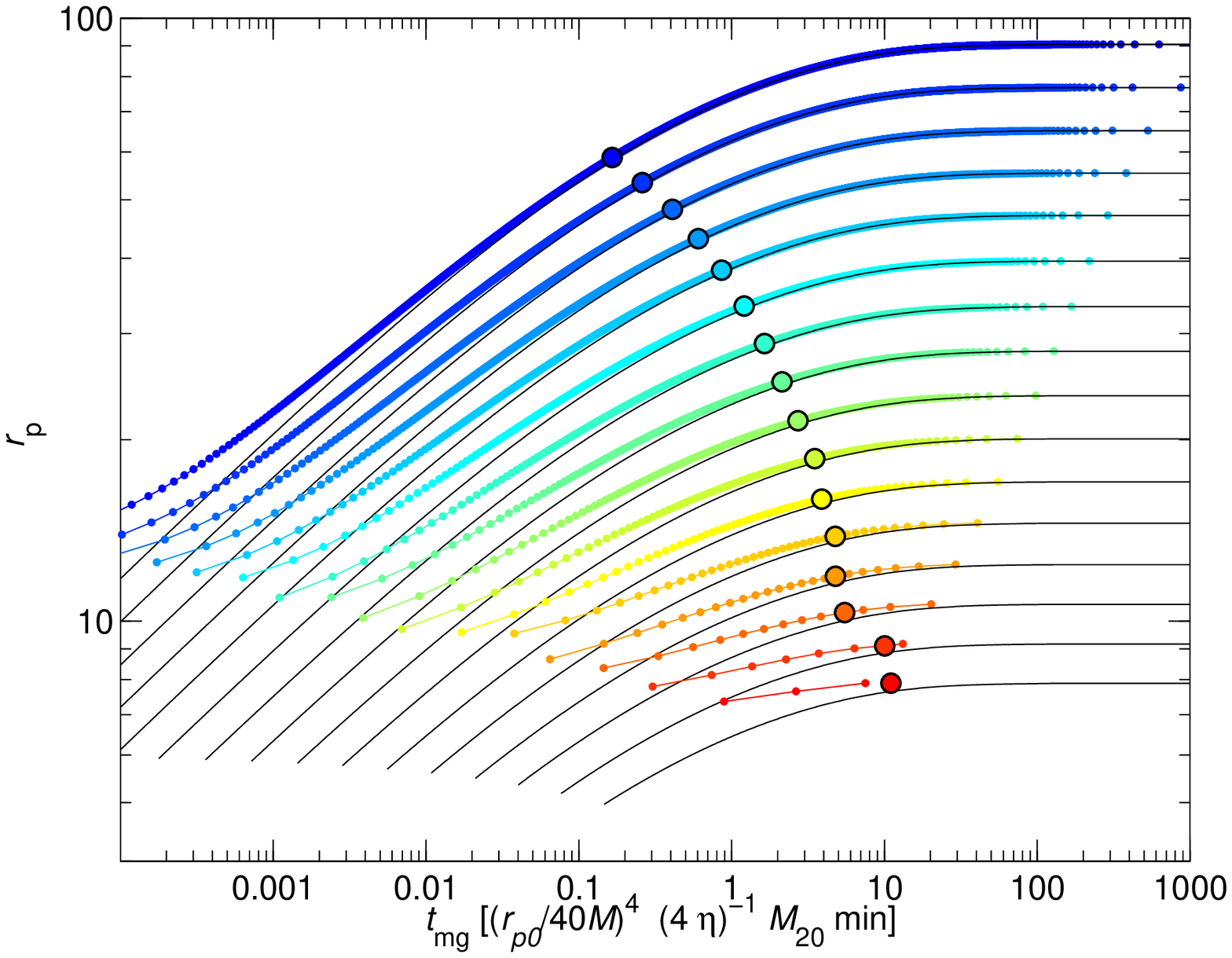}}
\mbox{\includegraphics[width=8.5cm]{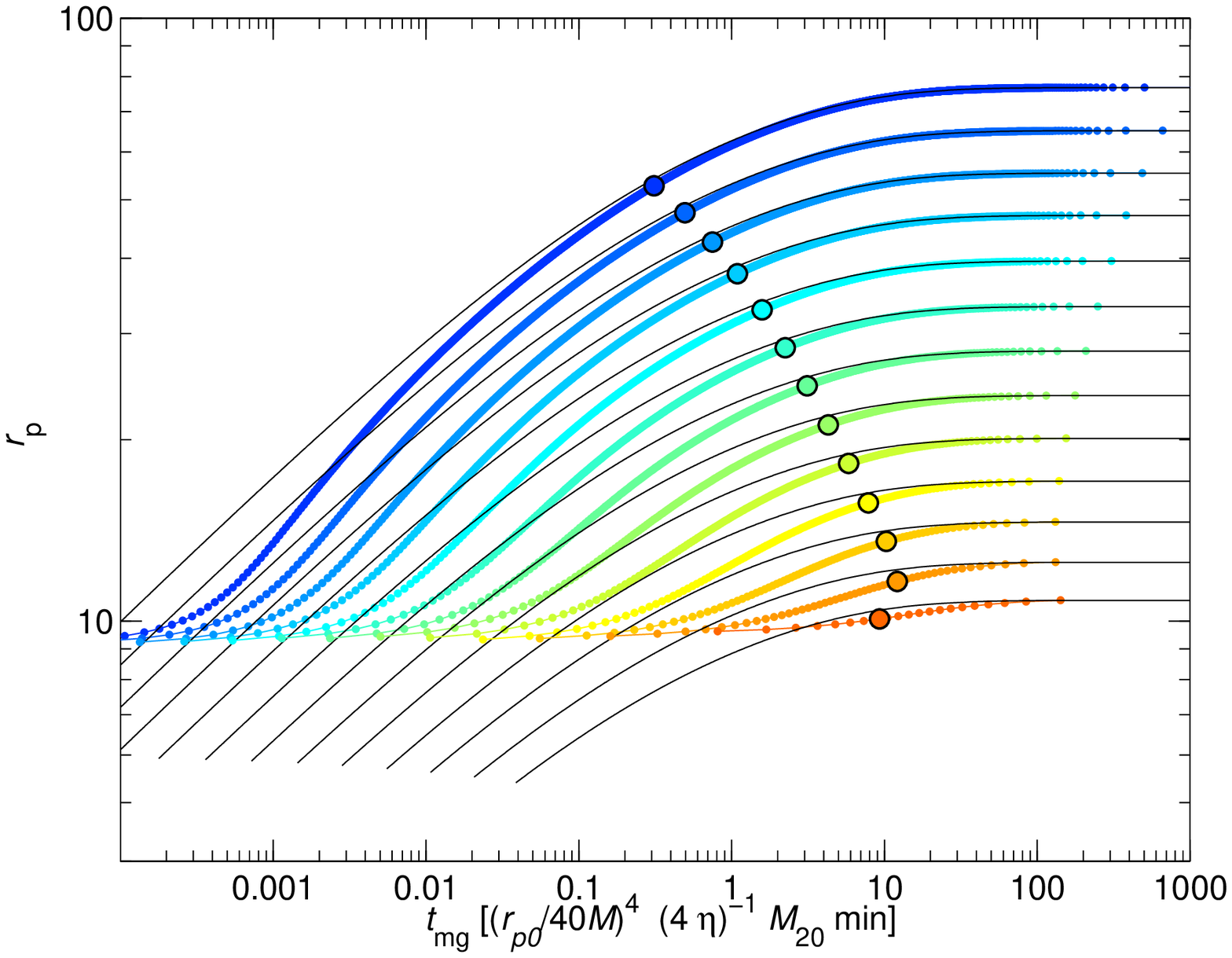}}\\
\mbox{\includegraphics[width=8.5cm]{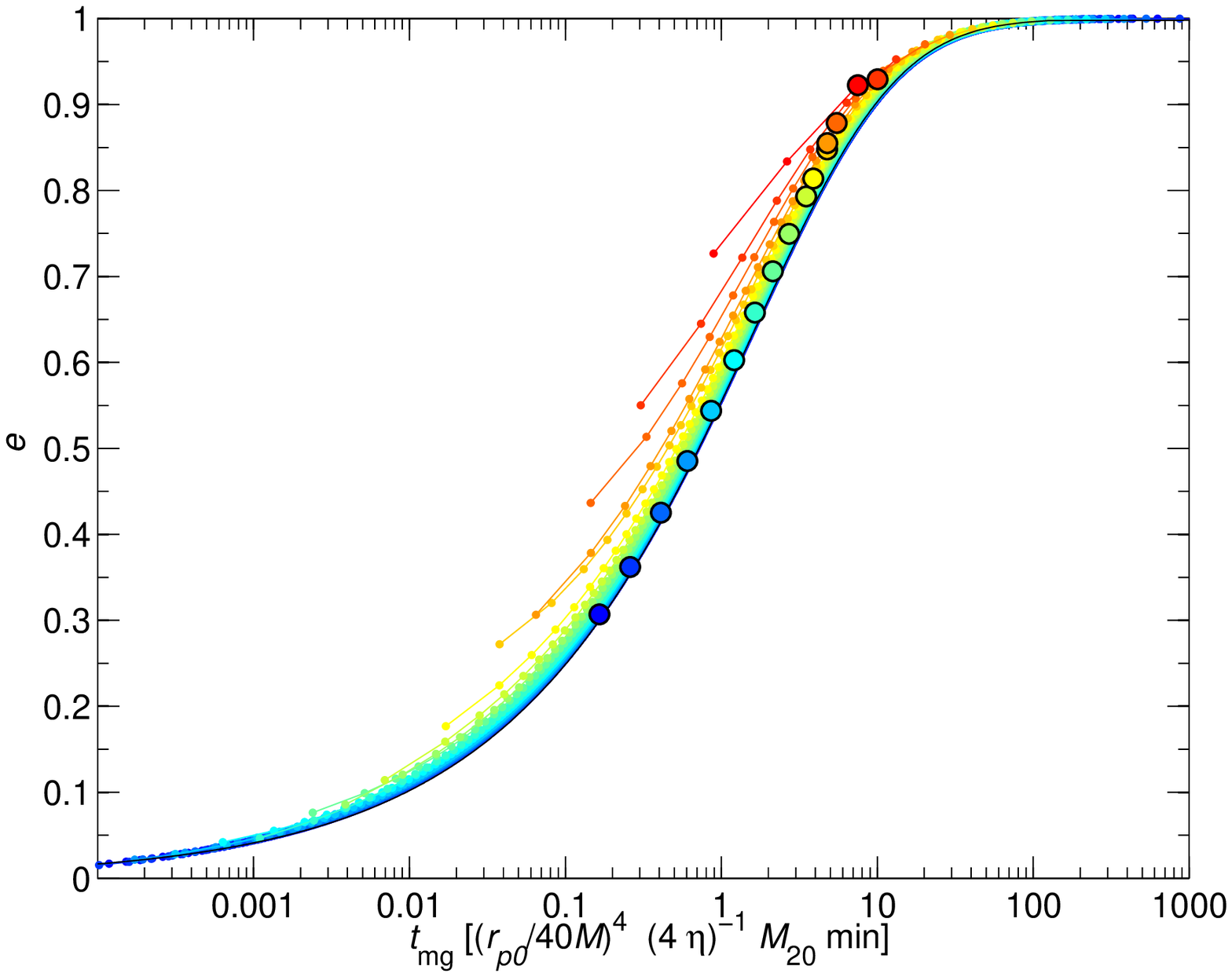}}
\mbox{\includegraphics[width=8.5cm]{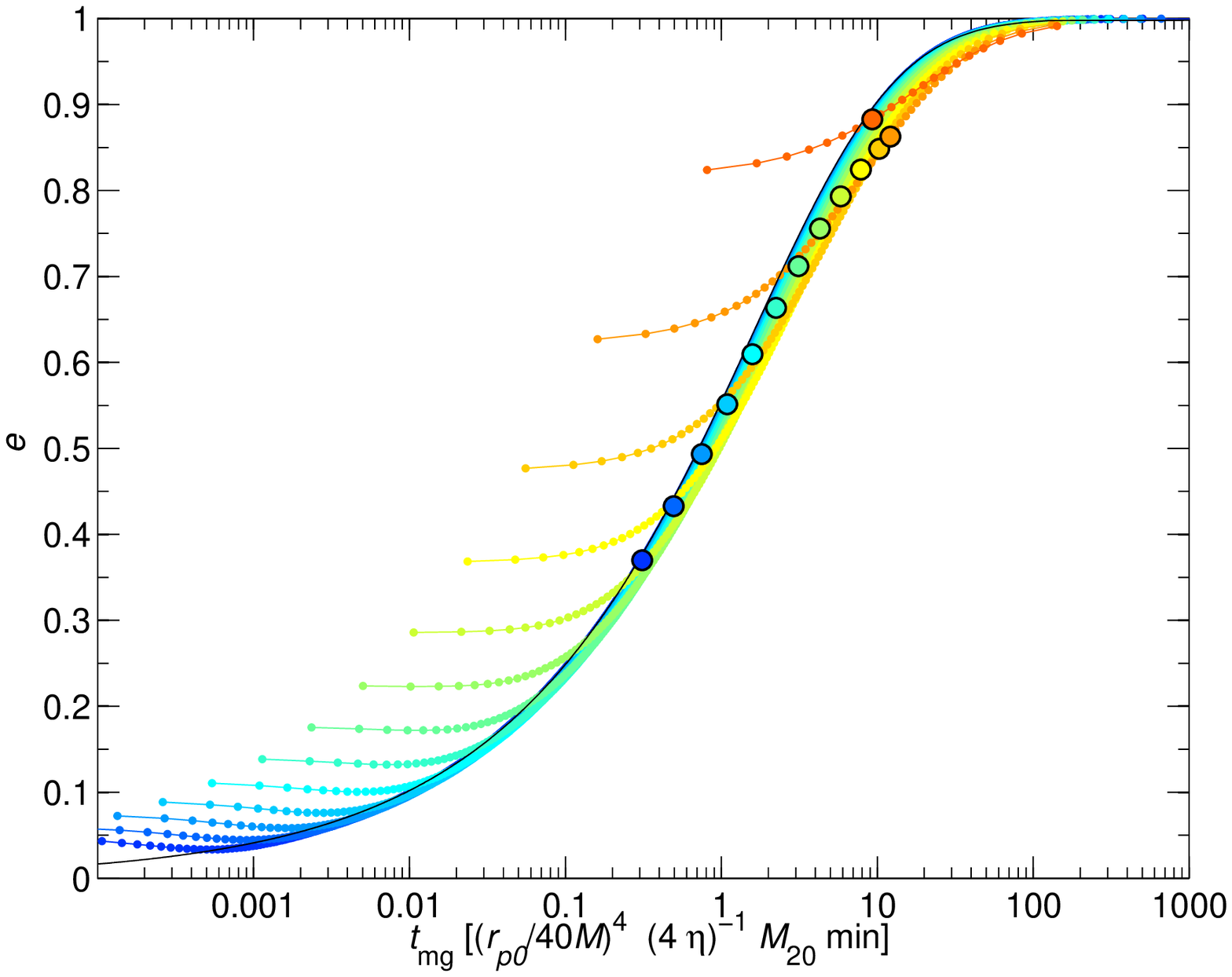}}
}
\caption{\label{f:eccentricity-time}
The evolution of $r_p$ and $e$ as a function of time to merger, for mass ratio $q=1$, zero spin, using
2.5PN (left) and 3.5PN (right). The black lines are the analytic solutions of Peters \cite{1964PhRv..136.1224P}.
Time is measured backwards from the innermost orbit of the simulation. Circles denote the transition
from the RB to the final chirp phase for $m_1=m_2=10\Msun$.
}
\end{figure*}

\begin{figure*}
\centering{
\mbox{\includegraphics[width=8.5cm]{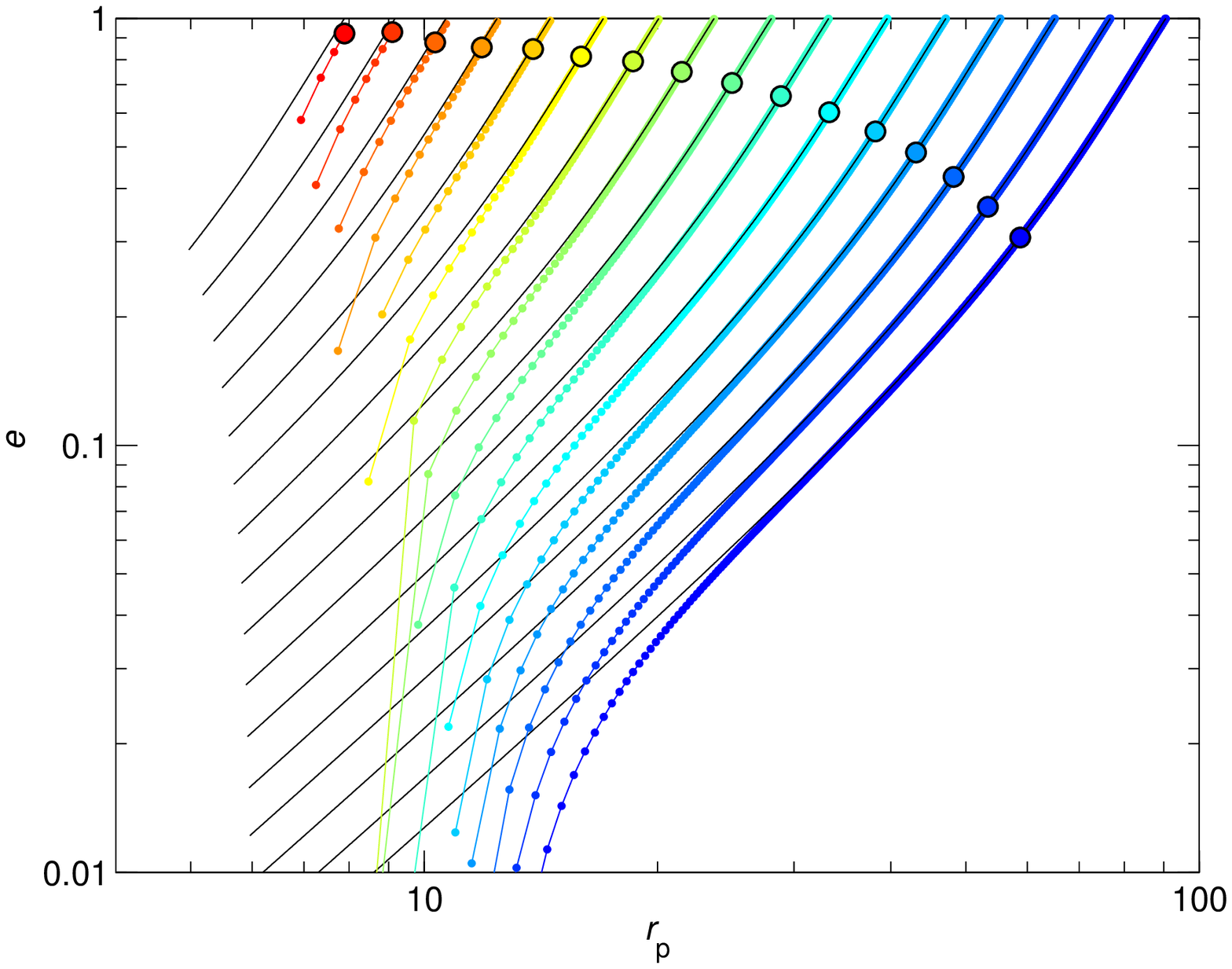}
\includegraphics[width=8.5cm]{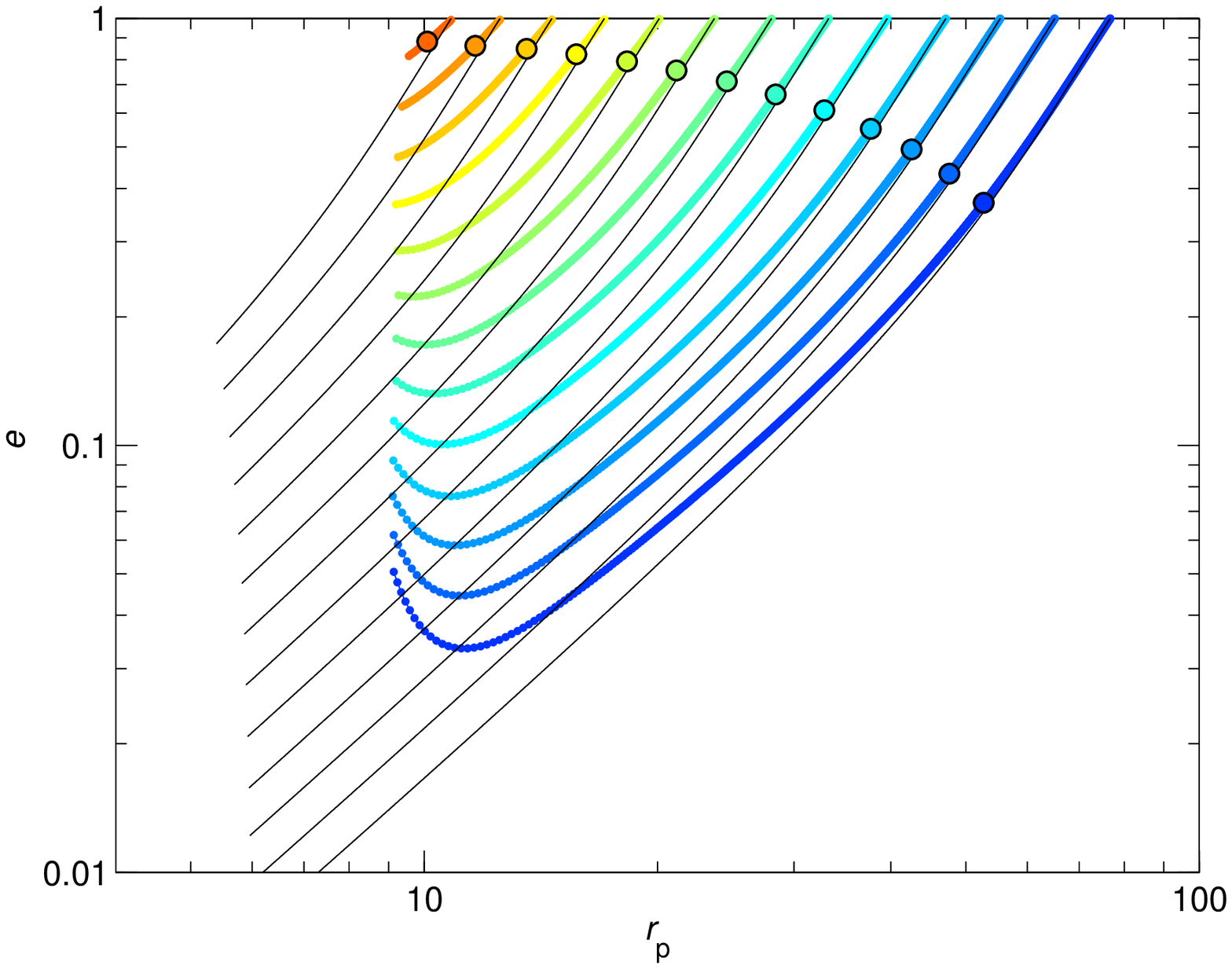}}
}
\caption{\label{f:orbits}
The evolution of eccentricity as a function of pericenter distance, for mass ratio $q=1$, zero spin, using
2.5PN (left) and 3.5PN (right). The black lines are the analytic solutions of Peters \cite{1964PhRv..136.1224P}.
}
\end{figure*}

Shortly after formation, the binary is very eccentric, the orbits are nearly radial. Due to GW losses near pericenter passage,
the apastron $r_a$ decreases faster than the pericenter distance, leading to a decrease in eccentricity.
The orbits exhibit large relativistic pericenter precession
(see Levin, McWilliams \& Contreras \cite{2011CQGra..28q5001L} for a gallery of orbital trajectories).

We compare our PN simulations with the leading order
analytical formulas of Peters \cite{1964PhRv..136.1224P} in Fig.~\ref{f:eccentricity-time} and \ref{f:orbits}.
In OKL, it was demonstrated that for the later, assuming an
initially parabolic orbit, $e(r_p/r_{p 0})$ and $e(t/t_0)$ are universal functions, independent of masses and
the impact parameter. These physical parameters affect only the time (here $t$ is measured from merger)
and length scales $(t_0,r_{p 0})$.
The evolution might be expected to be only slighly different for initially hyperbolic
orbits, as long as the velocity at close approach is dominated by the gravitational binding energy, not the initial kinetic energy.
The value of the relative velocity at infinity sets the maximum initial pericenter distance for binary formation.
We examine the inaccuracies in the evolutionary curves due to relativistic corrections.

Figure~\ref{f:eccentricity-time} and \ref{f:orbits} plot the orbital evolution $r_p(t/t_0)$, $e(t/t_0)$, and $e(r_p/r_{p0})$
for 16 different initial $(b,w)$ drawn from the white basin
of Fig.\ \ref{fig:basin} with $w=10^{-3} c$. The leading order analytic results are
shown as black curves, while the different panels correspond to 2.5PN and
3.5PN calculations.  The figures show that both the 2.5PN and
3.5PN calculation asymptote to the leading order eccentricity curve for large $r_p$. For widely formed binaries, the
deviations become significant interior to $r_p \sim 20 M$  in the 2.5PN calculation, or interior to $r_p \sim 40 M$ for the 3.5PN calculation,
while they are consistent to smaller separations if the initial $r_{p0}$ is less.
Part of the discrepancy between the simulated and the black analytic curves in Fig.~\ref{f:eccentricity-time}
is that in the latter case time is measured from merger, while in the former it is measured from the last simulated orbit.
This overall time shift is more prominent closer to merger on a logarithmic scale.
Interestingly, the full 2.5PN calculation decreases the eccentricity steeper as a function of pericenter distance than in the leading order
orbit-averaged approximation of OKL, while the 3.5PN calculations is
just the opposite, leading to a shallower eccentricity decrease.
This is consistent with the results that radiation reaction is
over-estimated at 2.5PN order and this over-estimate is tempered at 3.5PN
order \cite{2011CQGra..28q5001L}.
These two approximations bracket that used in OKL.
For higher $w$, corresponding to the innermost regions of galactic nuclei,
the highest initial $r_p$ values in the figures (dark blue curves) would not form binaries: $r_{p0,\max}\approx (93,50,25)$ for $w=(0.001,0.003,0.01)\,c$, respectively,
(see Eq.~(18) in OKL), but other curves with smaller $r_{p0}$ remain similar for different $w$.

Circles in Figs.~\ref{f:eccentricity-time} and \ref{f:orbits} mark the approximate boundary between the RB phase
and the final chirp, where the time duration between individual GW bursts (i.e. the orbital time) $\Delta t$ satisfies
$\Delta t \gtrsim 5/ f_{\min}$, where
$f_{\min}$ is the minimum frequency for a given detector. For Advanced LIGO, $f_{\min}\sim 10 \Hz$, so we require
$\Delta t \gtrsim 0.5\,$s in the RB phase. Note, that from Kepler's law, this amounts to an approximate constraint
on the semimajor axis, $a=r_p/(1-e) \gtrsim 87 M_{20}^{-2/3} M$, where $M$ is the total binary mass, and $M_{20}=M/20\Msun$.

Alternatively, we will also examine the signal to be in the RB phase if the GW signal is comprised of short duration
bursts and longer silent periods, requiring that the silent periods are at least a given factor (e.g. 2 or 4)
larger than the burst duration.
This later definition is equivalent to setting the eccentricity to be larger than some $e_{\min}$ in the RB phase (e.g.
$e_{\min}=0.45$ or 0.6), independent of the semimajor axis. In this case the end of the RB phase is a
horizontal line in Fig.~\ref{f:orbits}.

Simulations with different mass ratios and spins lead to similar curves as those in Fig.~\ref{f:orbits}. The 2.5PN and 3.5PN simulations
with mass ratio $q=0.1$ track the analytical leading order curves for $r_p(e)$ to somewhat smaller pericenter distances
(down to $r_p=(6, 7, 10, 15)M$ for $r_{p 0}=(8,10,20,40)$, respectively). The orbit evolves to much smaller eccentricites before plunging.
We have also run calculations with extremal spins in aligned, antialigned, and perpendicular configurations with respect to the
orbital angular momentum. The result is qualitatively very similar to the nonspinning case. In the aligned configuration, the time evolution
of the eccentricity tracks the Newtonian result much more closely than in the nonspinning case, while the spin-orbit precession adds
a slow periodic modulation to the evolution if the spins are initially perpendicularly oriented.

The 3.5PN approximation leads to an apparent eccentricity increase at $r_p\sim 10 M$ if $e\lesssim 0.15$.
This feature is close to the point at which we truncate the simulation
due to the breakdown of the approximation, so we take it with a
grain of salt
\cite{2011CQGra..28q5001L}. Essentially we are seeing the apastron
$r_a$ decrease more slowly than periastron $r_p$ leading to an
increase in $e=(r_a-r_p)/(r_a+r_p)$. For a dissipating orbit,
eccentricity is not a precisely defined quantity, but the
GW signal does show qualitative measures of this $e$,
such as a broadband character, noticeable for $e \gtrsim 0.1$.
In the extreme mass ratio case, eccentricity increase is known to occur only for
marginally plunging orbits with $r_{p 0}\sim 4$
\cite{1994PhRvD..50.3816C}.
The 2.5PN and 3.5PN calculations agree until the end of the RB phase marked by circles
in Figs.~\ref{f:eccentricity-time} and \ref{f:orbits}. The calculations are roughly consistent
for the highly eccentric orbits $e\gtrsim 0.6$, except for small $r_{p0}\lesssim 15\,M$ which represent zoom-whirl orbits.
In general, the true orbits are expected to whirl at small separations
and exhibit larger precession than in the PN calculations.

\section{Gravitational waves and their detection}

We obtain the emitted GWs during the orbital evolution as a function of time.
In practice, we calculate the instantaneous $h_{+}(t)$ and $h_{\times}(t)$ polarizations
of the strain amplitude in the direction of the orbital axis
from the instantaneous phase space elements (for details, see \cite{2011CQGra..28q5001L}).
For each mass ratio, we can use a single simulation to describe sources with arbitrary total masses and source distances,
by scaling the amplitude and  time proportionally with $M_z/\dL$  and $M_z$, respectively, where $M_z = (1+z) M$
is the cosmological redshifted total mass and $\dL$ is the luminosity distance.

Given $h_{+}(t)$ and $h_{\times}(t)$, the instrument measures a combination
\begin{equation}
 h(t) = F_{+} h_{+}(t) + F_{\times} h_{\times}(t)
\end{equation}
where $F_+$ and $F_{\times}$ are the antenna beam pattern coefficients, which depend on the orientation of the detector
with respect to the binary (see Eq.~(104) in Ref.~\cite{1987thyg.book..330T}), and satisfy $0\leq |F_{+,\times}| \leq 1$, where
$F_{+}=1$ if $F_{\times}=0$ (and vice versa), and on average $\langle F_{+,\times}^2\rangle=1/5$.
We discuss how we make inferences for an average binary orientation from a waveform corresponding
to the optimal orientation in Appendix~\ref{app:angular}.

The simulations confirm the qualitative waveform features presented in OKL. During the first part of the evolution, when the eccentricity is very large,
the binary emits repeated GW bursts (RB) during successive close approaches. The relative amplitude of the two polarizations are modulated by
GR precession. The time separation between successive bursts decreases rapidly on a logarithmic scale, as the eccentricity decreases. The time
duration of individual bursts changes much more slowly. Eventually, as the eccentricity becomes small, the RB phase evolves toward
a continuous chirp signal.

In order to assess the detectability of the waveforms, we calculate the numerical FFT of the waveform
for both polarizations, $\tilde h_{+,\times}(f)$, and compare to the sensitivity level of GW instruments.
This is more accurate than the OKL estimate, as that relied on the
stationary phase approximation to estimate the Fourier amplitude of each harmonic.

The SNR is given by
\begin{equation}\label{e:snr}
\left\langle\frac{S}{N}\right\rangle^2 =
4 \int_{f_{\min}}^{f_{\max}} \frac{ \tilde{h}^2(f)}{S_n(f)} \D f = \int_{f_{\min}}^{f_{\max}} \left[\frac{ 2f\tilde{h}(f)}{\sqrt{f S_n(f)}}\right]^2 \frac{\D f}{f}
\end{equation}
where $S_n(f)$ is the one sided spectral noise density and
$\tilde h(f)$ is the sky position, binary orientation, and polarization averaged GW signal spectral amplitude (see Appendix~\ref{app:angular}).

Eq.~(\ref{e:snr}) shows that $2 f \tilde h$
and $\sqrt{f S_n}$ correspond to the angular averaged spectral signal amplitude and RMS noise amplitude per logarithmic
frequency bin, whose ratio gives the SNR per logarithmic frequency bin. In the following we compare these
dimensionless quantities when discussing the detectability of the signal.

\subsection{Circular orbits}

\begin{figure}
\centering{
\mbox{\includegraphics[width=8.6cm]{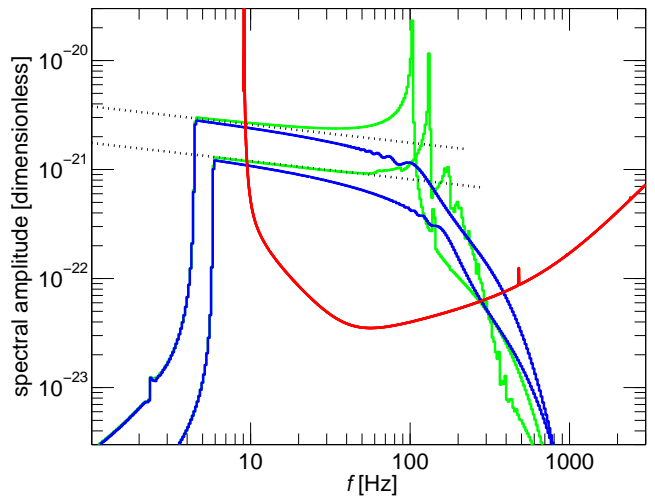}}}
\caption{\label{f:AdLIGO-spectra-circ}
The characteristic spectral amplitude for circular orbit (zero spin) initial conditions $r_0=80M$,
for 2.5PN (blue) and 3.5PN (green) and masses $m_1=m_2=10\Msun$ (top curves) and $m_1=0.1m_2=1.4\Msun$ (bottom curves).
Dotted lines show the analytical stationary phase approximation using the 2.5PN orbit-averaged flux of Poisson \& Will \cite{1995PhRvD..52..848P}.
}
\end{figure}

We first examine the convergence of 2.5PN and 3.5PN calculations for circular initial conditions.
Figure~\ref{f:AdLIGO-spectra-circ} shows the GW spectra in this case, upper and lower curves correspond
to BH/BH and BH/NS binaries. The dotted lines show the analytical spectra for 2.5PN orbital-averaged flux calculation for circular orbits
with the stationary phase approximation (SPA), see Eq.~(\ref{e:SPA}) in Appendix~\ref{app:angular} below.
Both the 2.5PN and 3.5PN calculations asymptote the orbit-averaged flux spectra for large separations or small frequencies,
but at higher frequencies, they lead to systematically lower and higher GW spectral amplitudes, respectively. Interestingly,
the orbit-averaged flux result is well off our 2.5PN spectral amplitude at $f\lesssim 50 \Hz$
(i.e. $r \lesssim 20 M$ for BH/NS binaries), but it is very close to our 3.5PN calculation.
The very strong final peak and the final upper harmonics are artifacts of the 3.5PN calculation,
as the orbital evolution slows down there around $10M$, where $f_{\rm cut}=10^{-3/2}\pi^{-1}M^{-1}=100\,\Hz\times (M/20\Msun)^{-1}$
and the eccentricity increases in the calculation (see Fig.~\ref{f:orbits}).
A similar spectral increase was identified by Buonanno et al. \cite{2003PhRvD..67b4016B} using a different approach,
where the GW flux was calculated to 1PN beyond leading order, corresponding to 3.5PN in our calculation, but it is not present
using higher order corrections to the flux.
Such higher order orbital averaged flux calculations are consistent with numerical simulations for quasicircular inspirals
\cite{2007PhRvD..76l4038B,2007PhRvL..99r1101B,2007PhRvD..75l4018B,2010PhRvD..82b4033H}.
In the following we present both the 2.5PN and 3.5PN calculation results for eccentric orbits
to gauge the error of our calculations in the RB phase.

\subsection{Time evolution of the GW spectra}

\begin{figure*}
\centering{
\mbox{
\includegraphics[width=6.3cm]{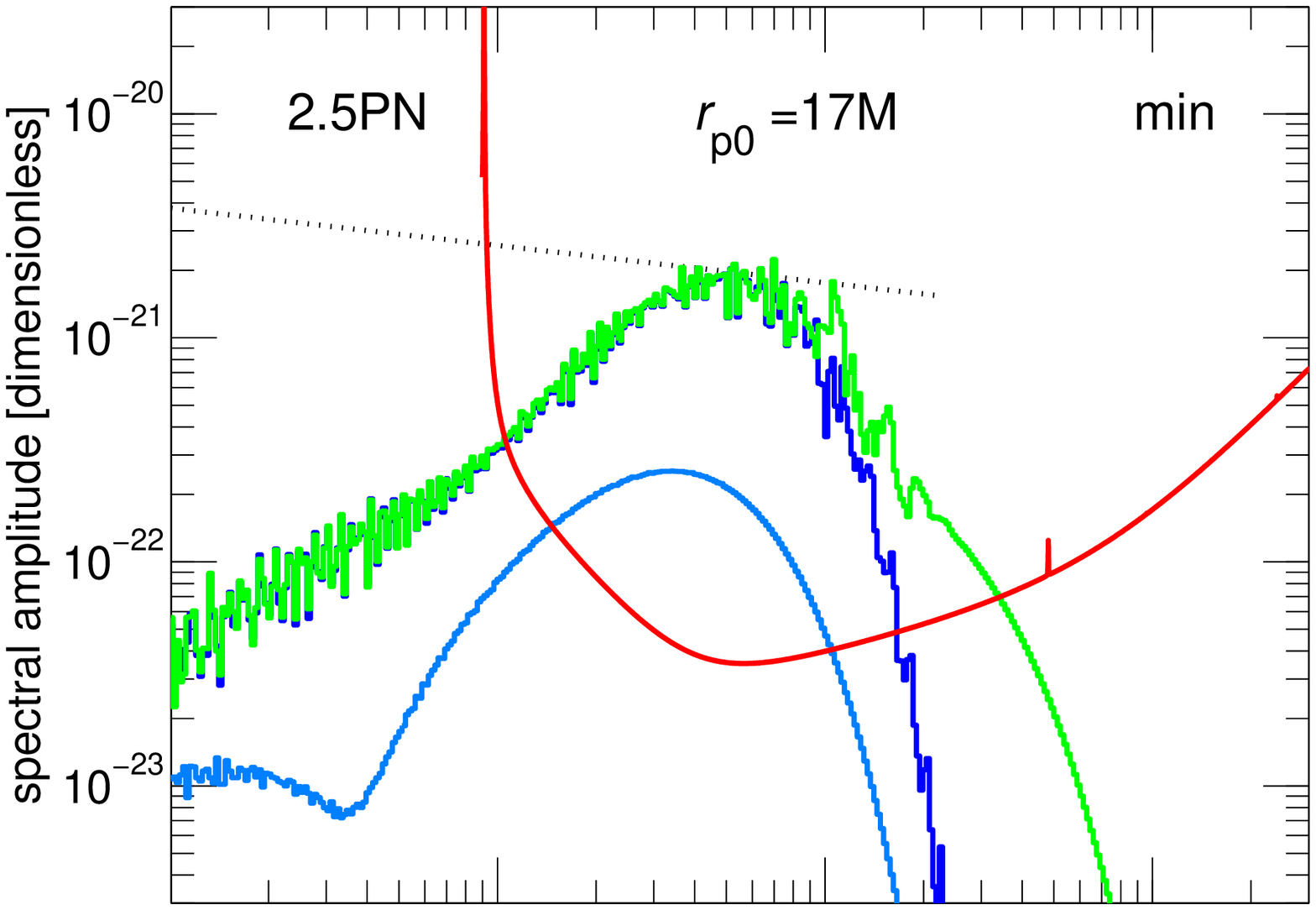}
\includegraphics[width=5.5cm]{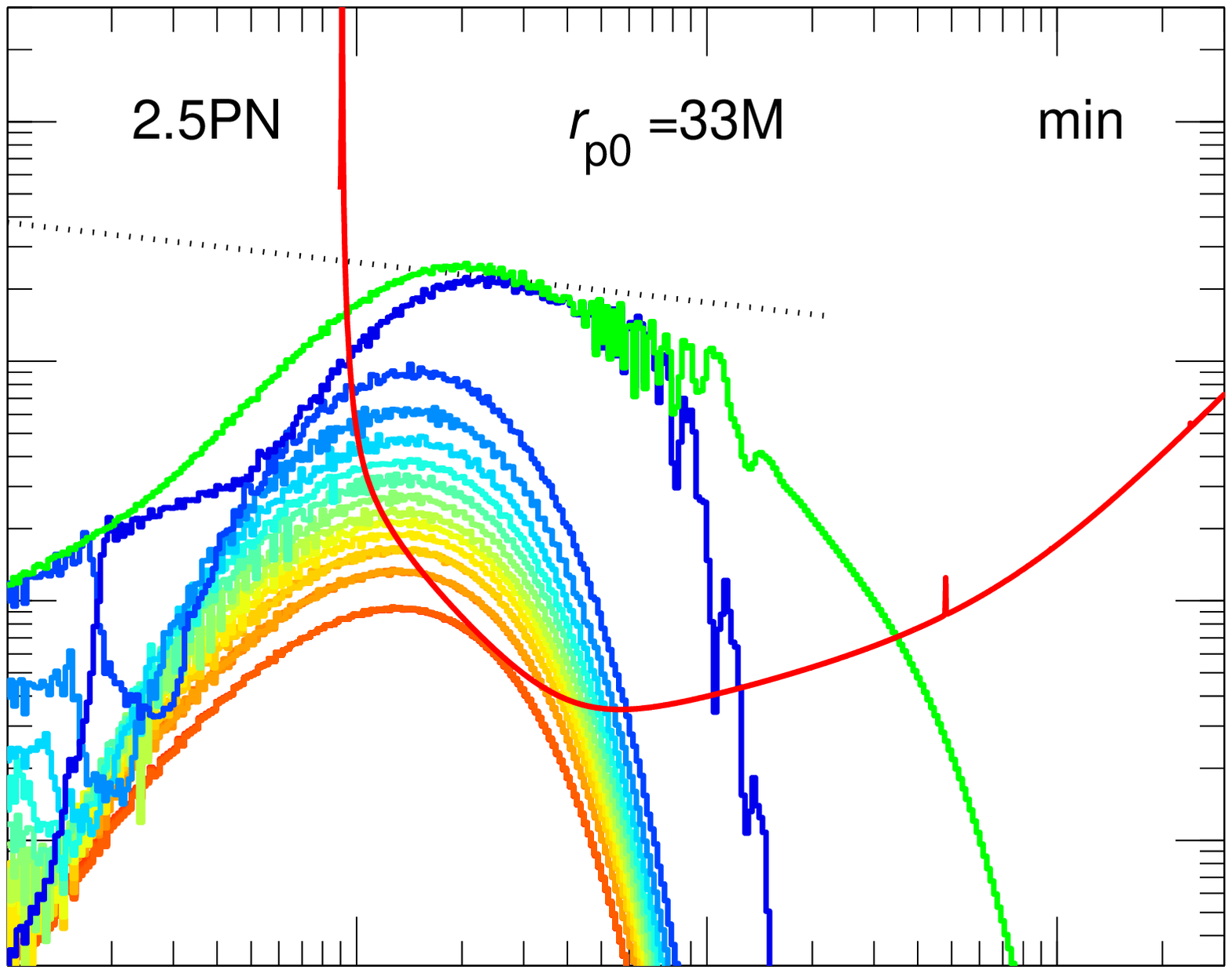}
\includegraphics[width=5.5cm]{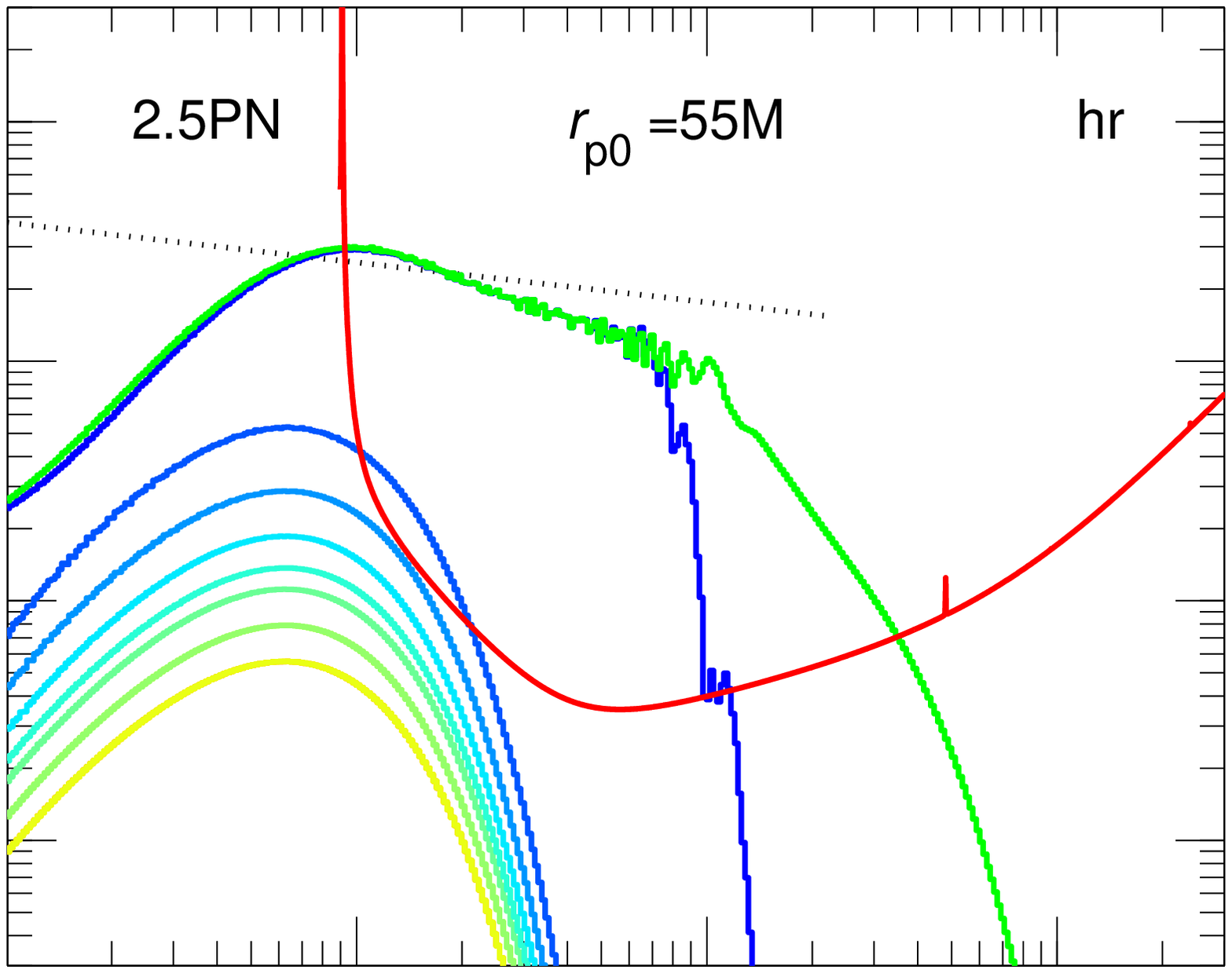}}\\
\mbox{
\;\includegraphics[width=6.3cm]{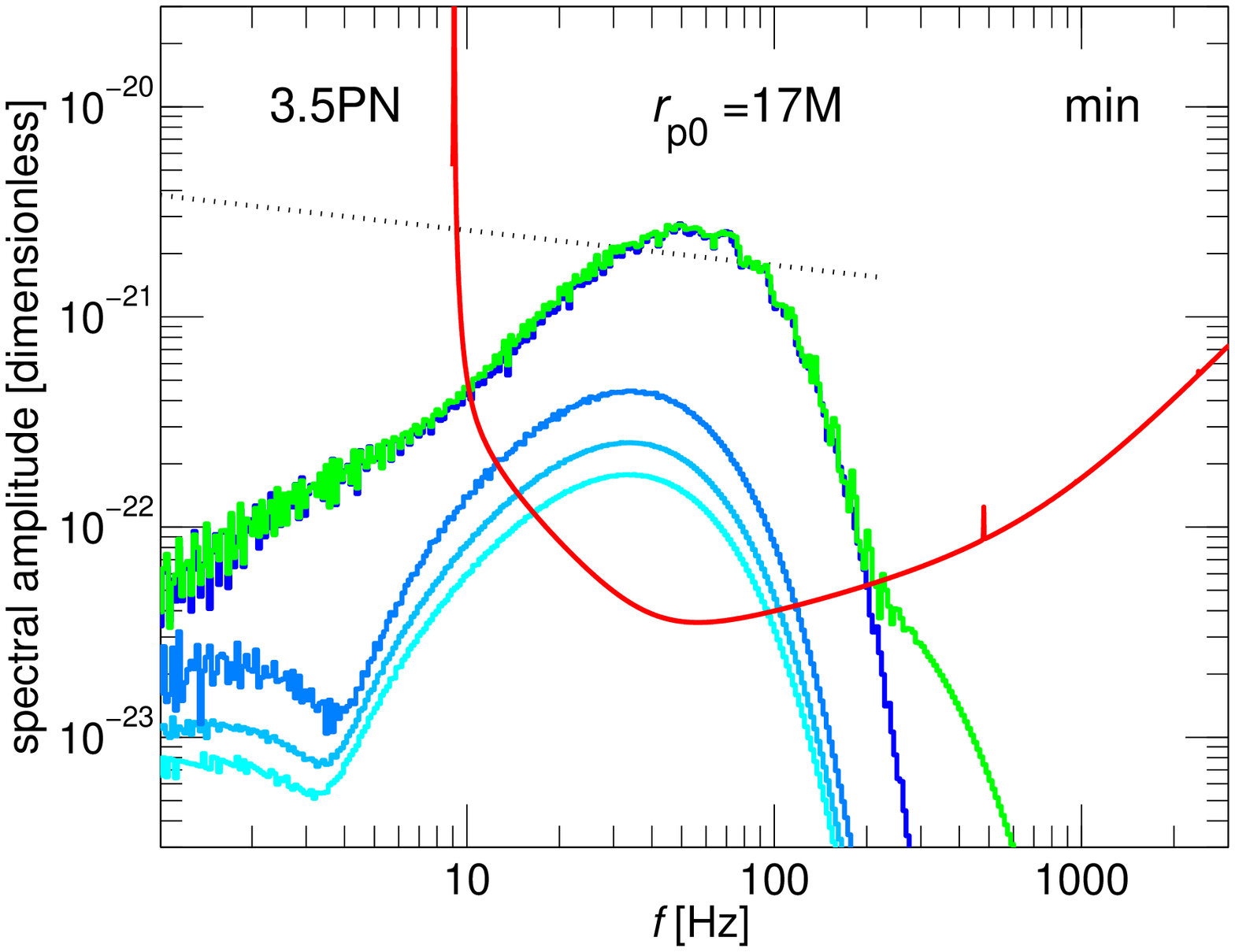}
\includegraphics[width=5.5cm]{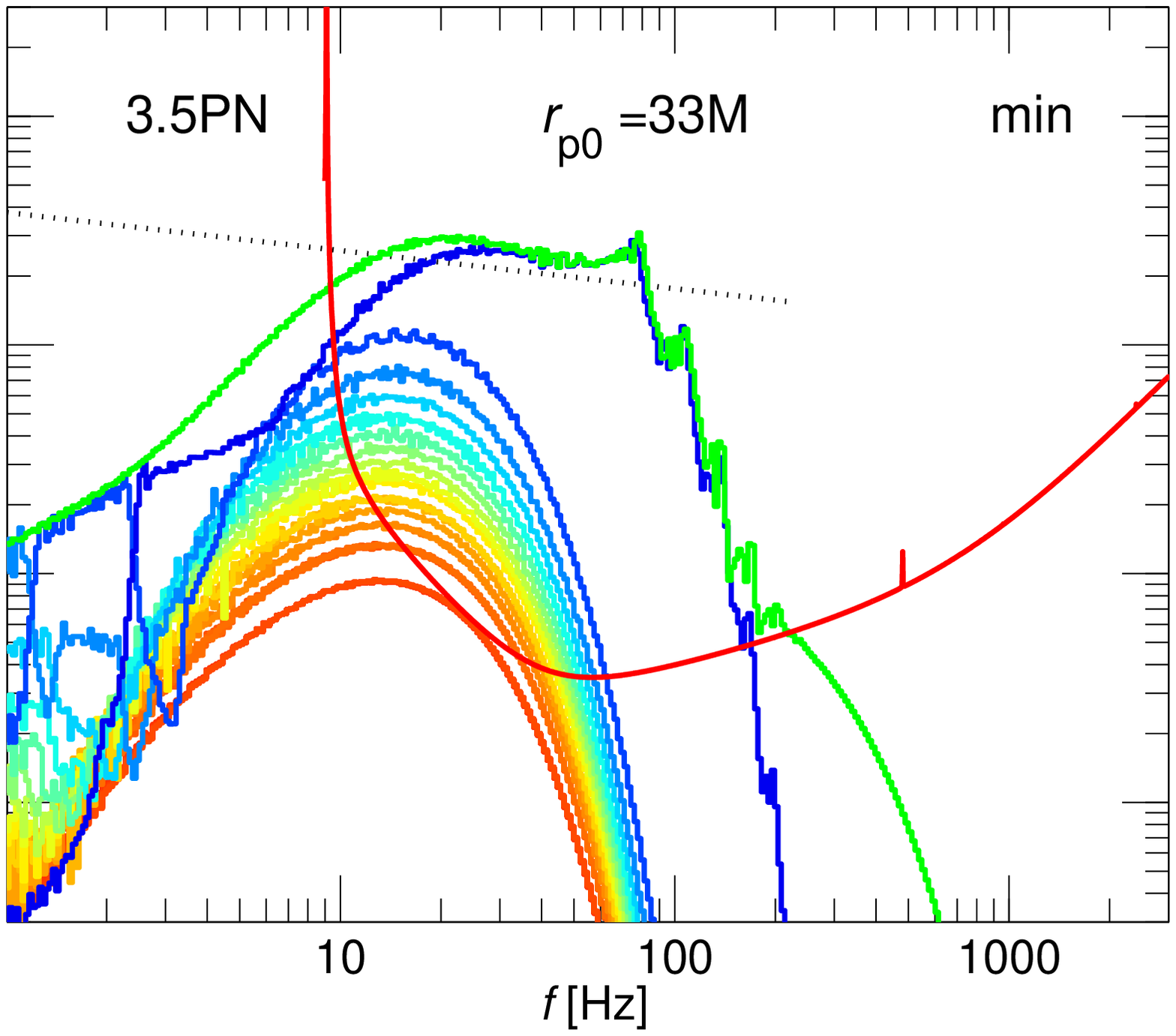}
\includegraphics[width=5.5cm]{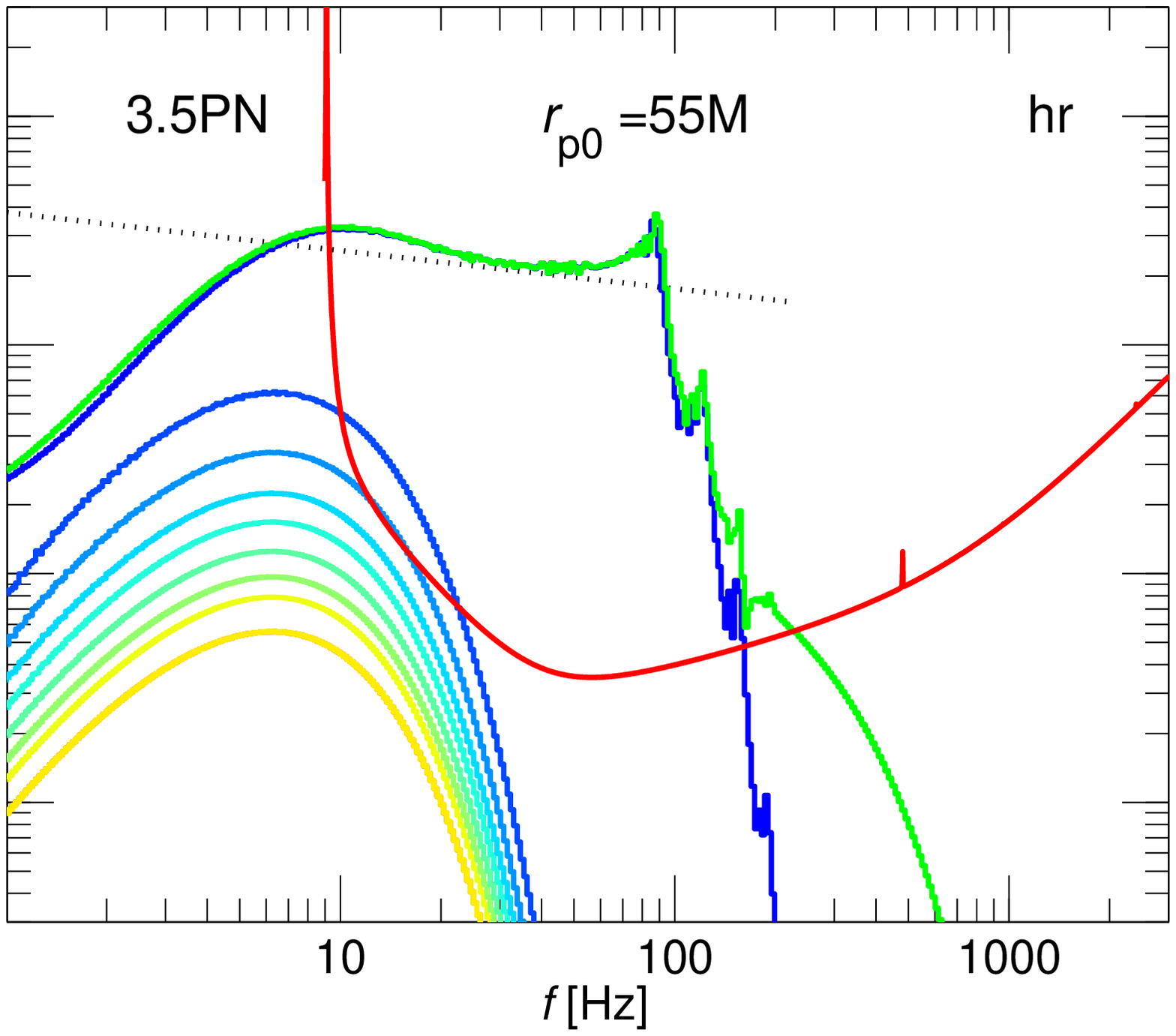}}
}
\caption{\label{f:AdLIGO-spectra-ecc1}
The characteristic spectral amplitude for GW capture orbits
for 2.5PN (panels in row 1) and 3.5PN (row 2) calculations, and masses $m_1=m_2=10\Msun$
for $r_{p0}=(17,33,55)M$ (left, middle, and right panels)
for a binary at $100\,$Mpc with an average orientation and zero spin.
Top green curves correspond to the full waveform, others show the contribution of 1 minute (left and middle)
and 1 hour segments (right panels), respectively. Thick red line is the Advanced LIGO sensitivity.
}
\end{figure*}
\begin{figure*}
\centering{
\mbox{
\includegraphics[width=6.3cm]{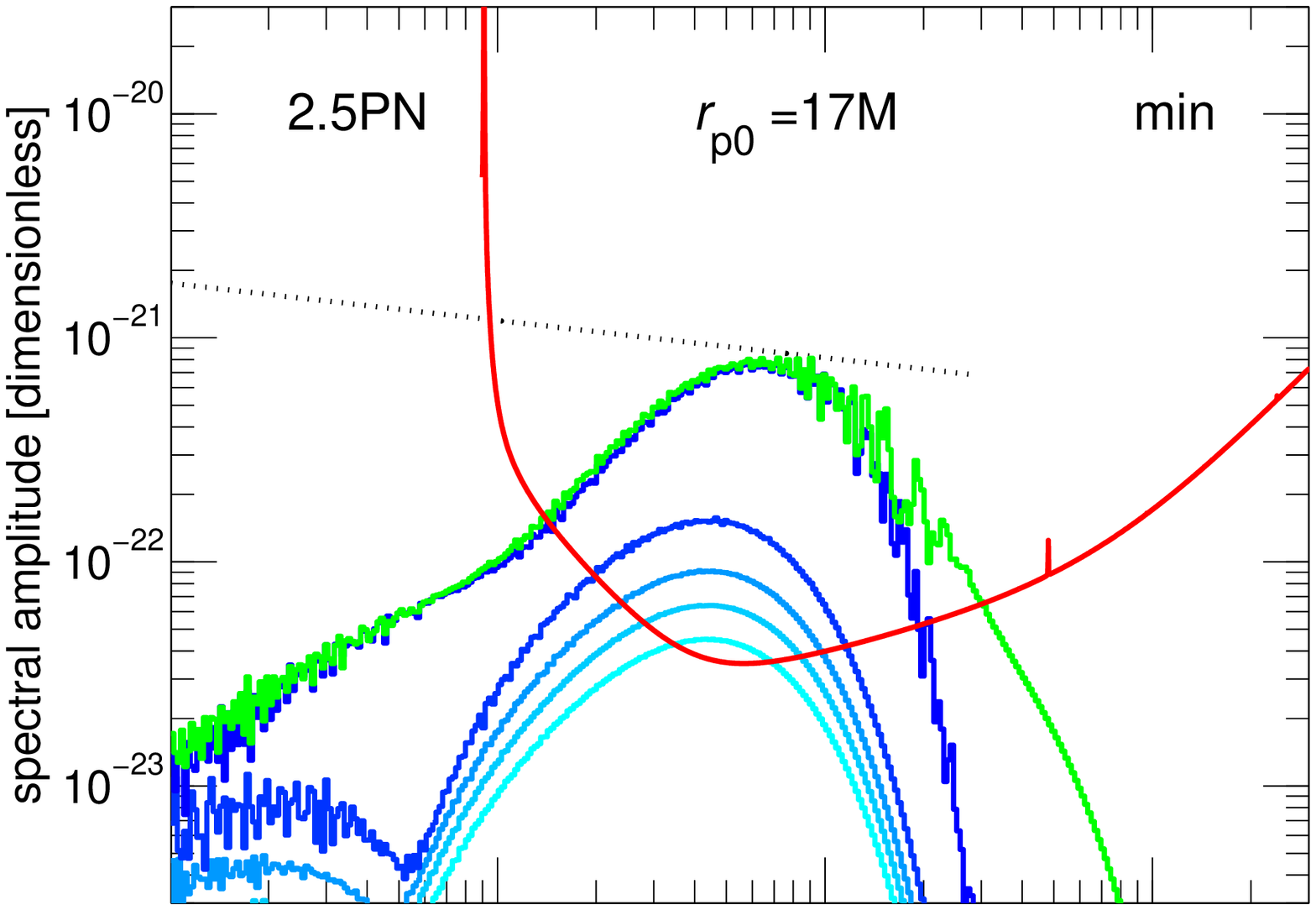}
\includegraphics[width=5.5cm]{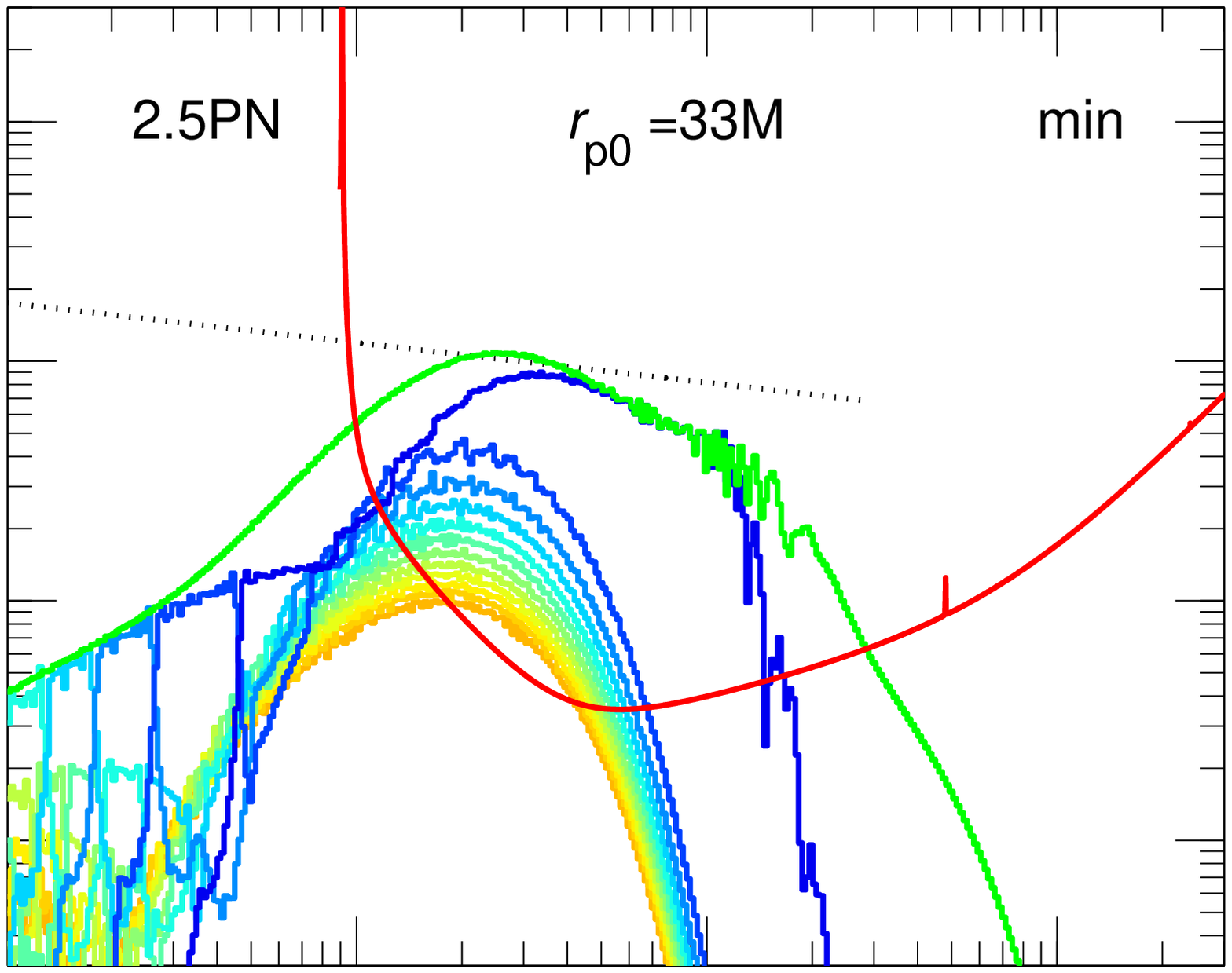}
\includegraphics[width=5.5cm]{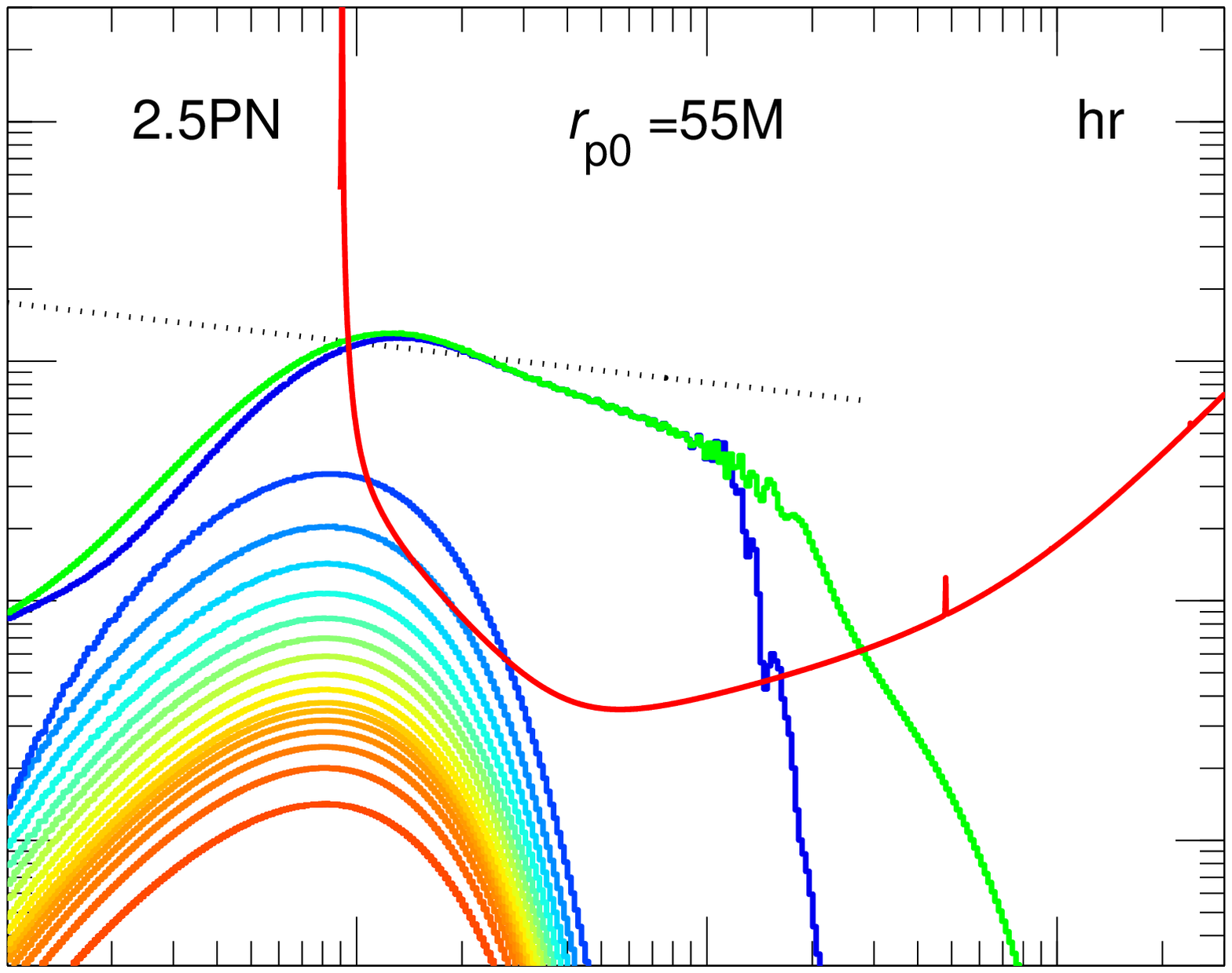}}\\
\mbox{
\;\includegraphics[width=6.3cm]{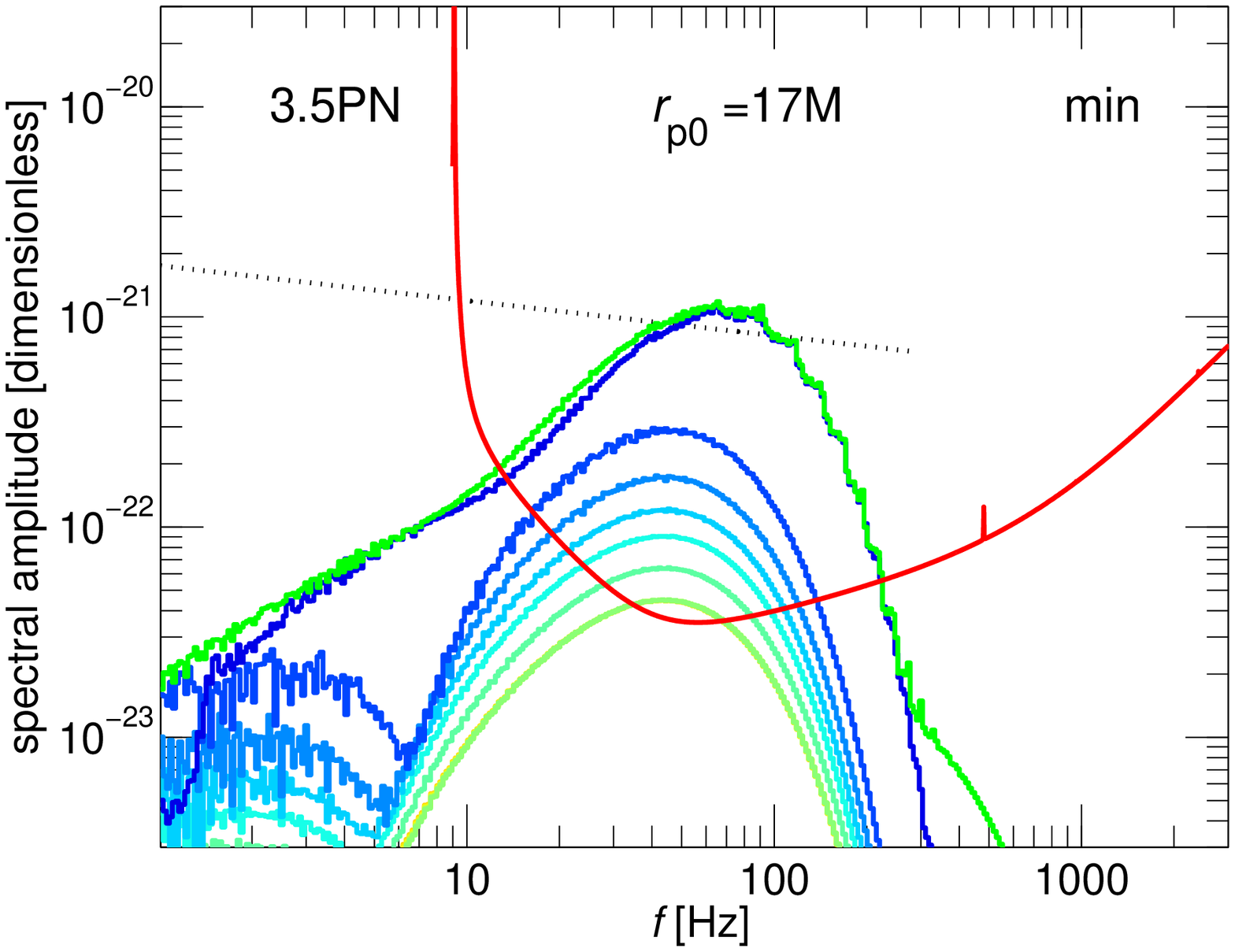}
\includegraphics[width=5.5cm]{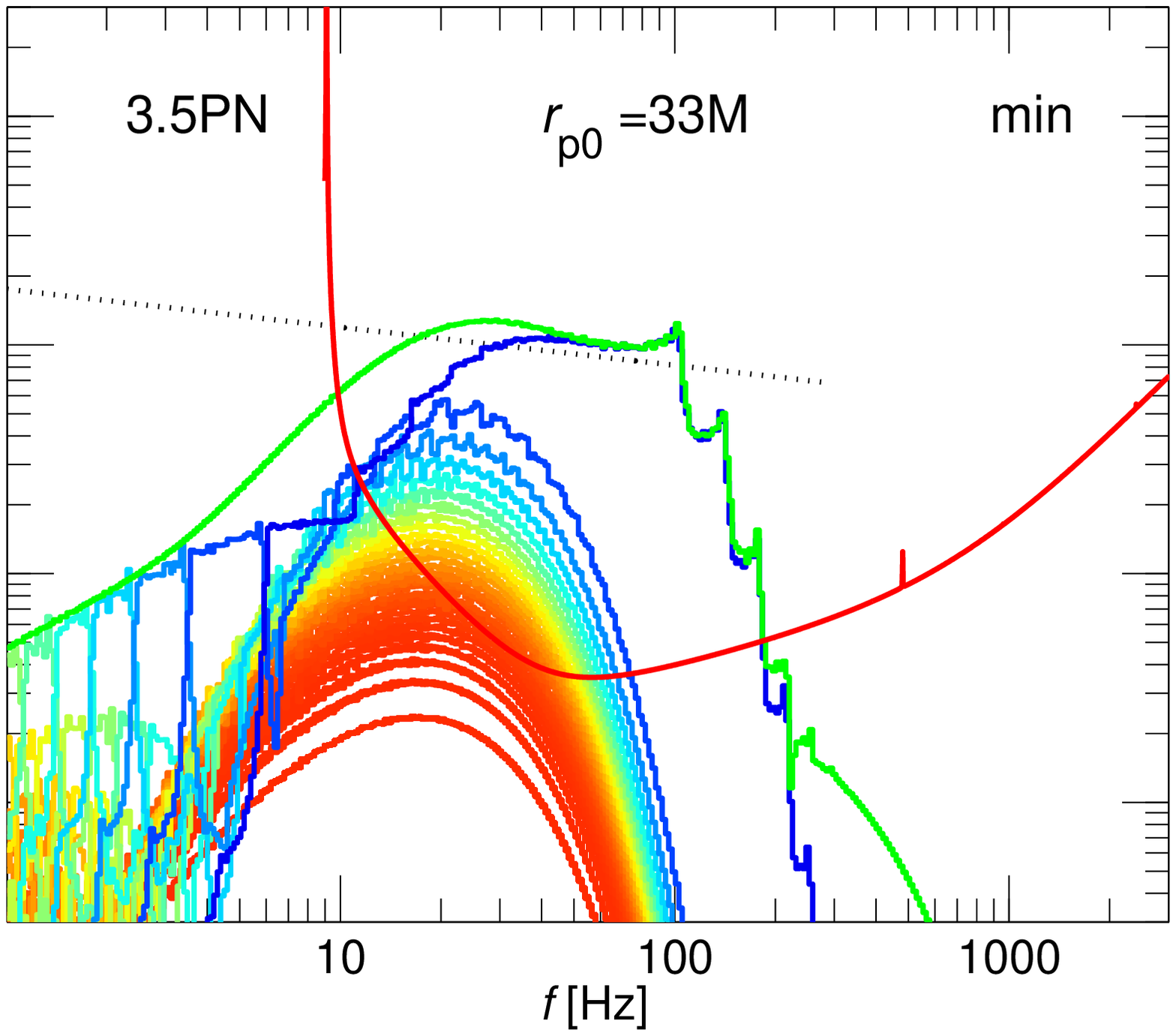}
\includegraphics[width=5.5cm]{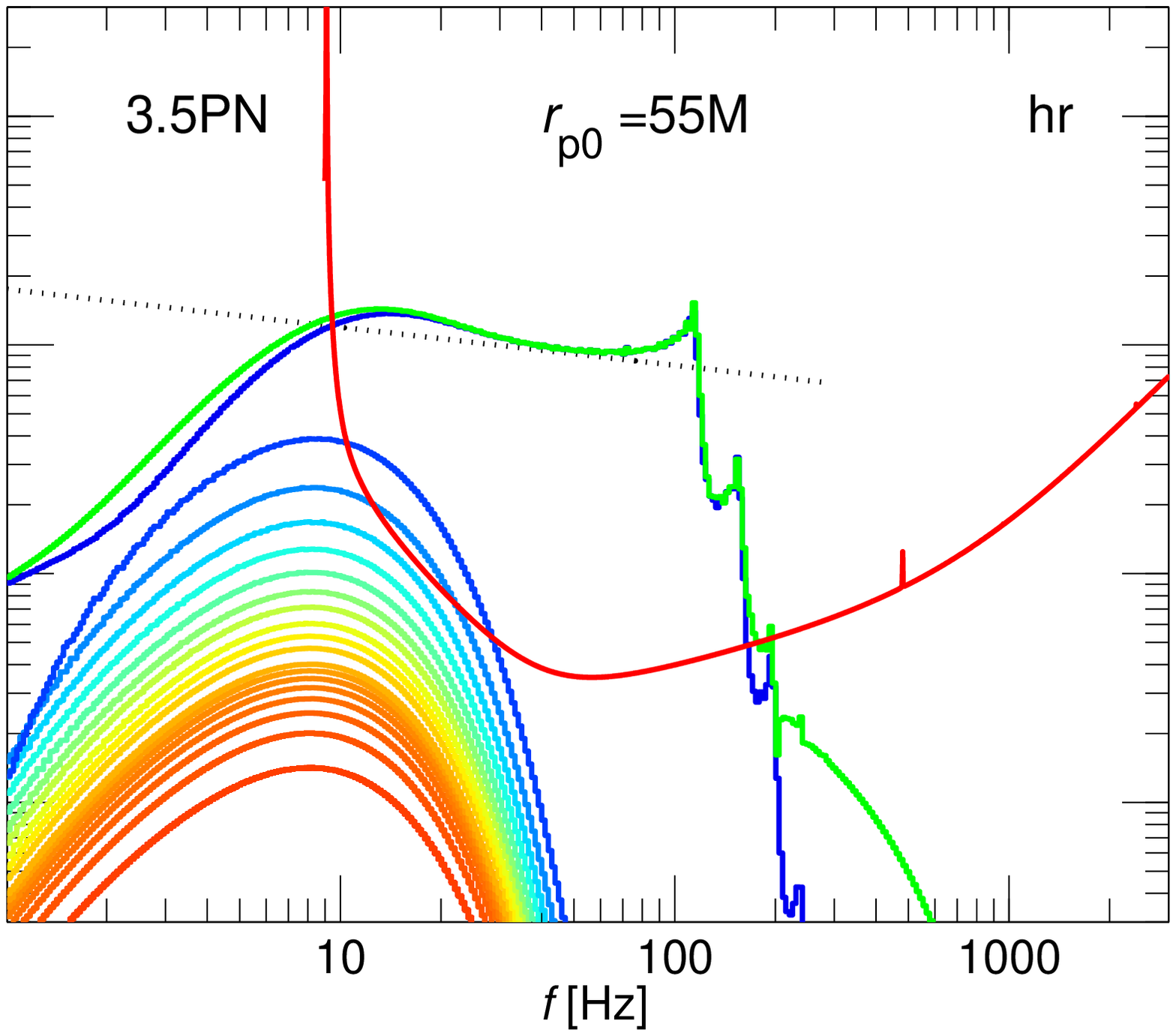}}
}
\caption{\label{f:AdLIGO-spectra-ecc2}
Same as Fig.~\ref{f:AdLIGO-spectra-ecc1} but for $m_1=0.1m_2=1.4\Msun$ resembling BH/NS encounters.
}
\end{figure*}

Figures~\ref{f:AdLIGO-spectra-ecc1} and \ref{f:AdLIGO-spectra-ecc2} show the angle-averaged
GW spectral amplitude, for different
initial pericenter distances $r_{p0}=(17,33,55)M$, using (2.5, 3.5)PN orders, and mass ratios $q=(1,0.1)$.
The colored curves represent separate minute and hour segments as marked,
while the top green curve shows the total root-sum-squared (RSS) spectra of the full waveform.
The dotted line shows the spectral amplitude for circular inspirals for reference (see Eq.~(\ref{e:SPA})).
The signal is in the  Advanced LIGO band and well above the sensitivity
level  for minutes to hours. Initially, in the RB phase, the waveforms are broadband
within a short-duration followed by long silent periods.
Although the orbital frequency $f_{\rm orb}$ is well outside the LIGO band
here\footnote{ In fact, depending on mass, $f_p$ and $f_{\rm orb}$ can be
in the LIGO and LISA band coincidentally in the RB phase, see Sec.~\ref{s:multiwavelength} below.},
the characteristic frequency of the bursts are the inverse timescale of pericenter passage,
$f_p = f_{\rm orb} (1+e)^{1/2}/(1-e)^{3/2} \gg f_{\rm orb}$.
Remarkably, the signal transitions from the RB
phase to a chirp signal in the detector band, as the eccentricity quickly decreases
in time.

\begin{figure*}
\centering{
\mbox{
\includegraphics[width=8.5cm]{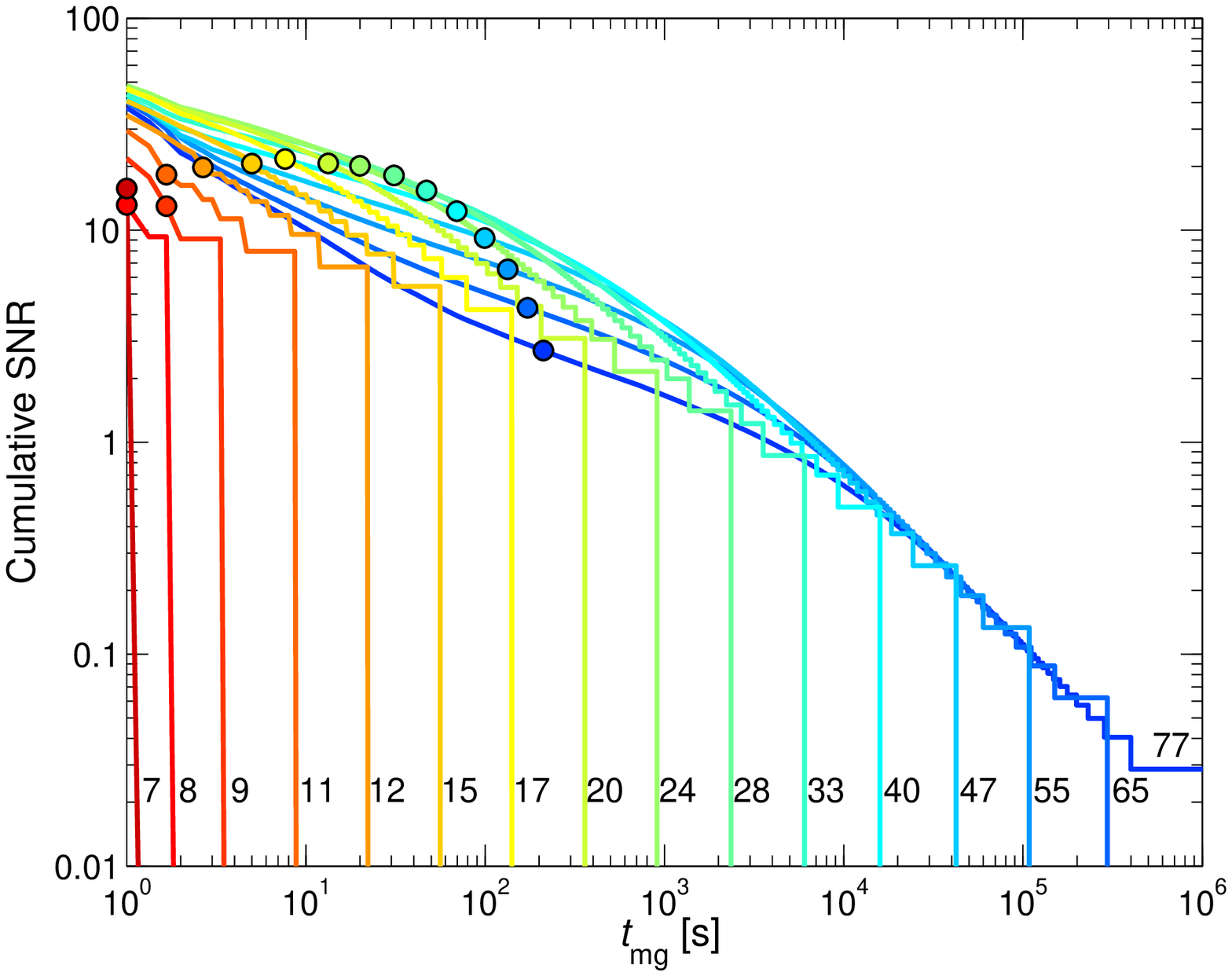}
\includegraphics[width=8.5cm]{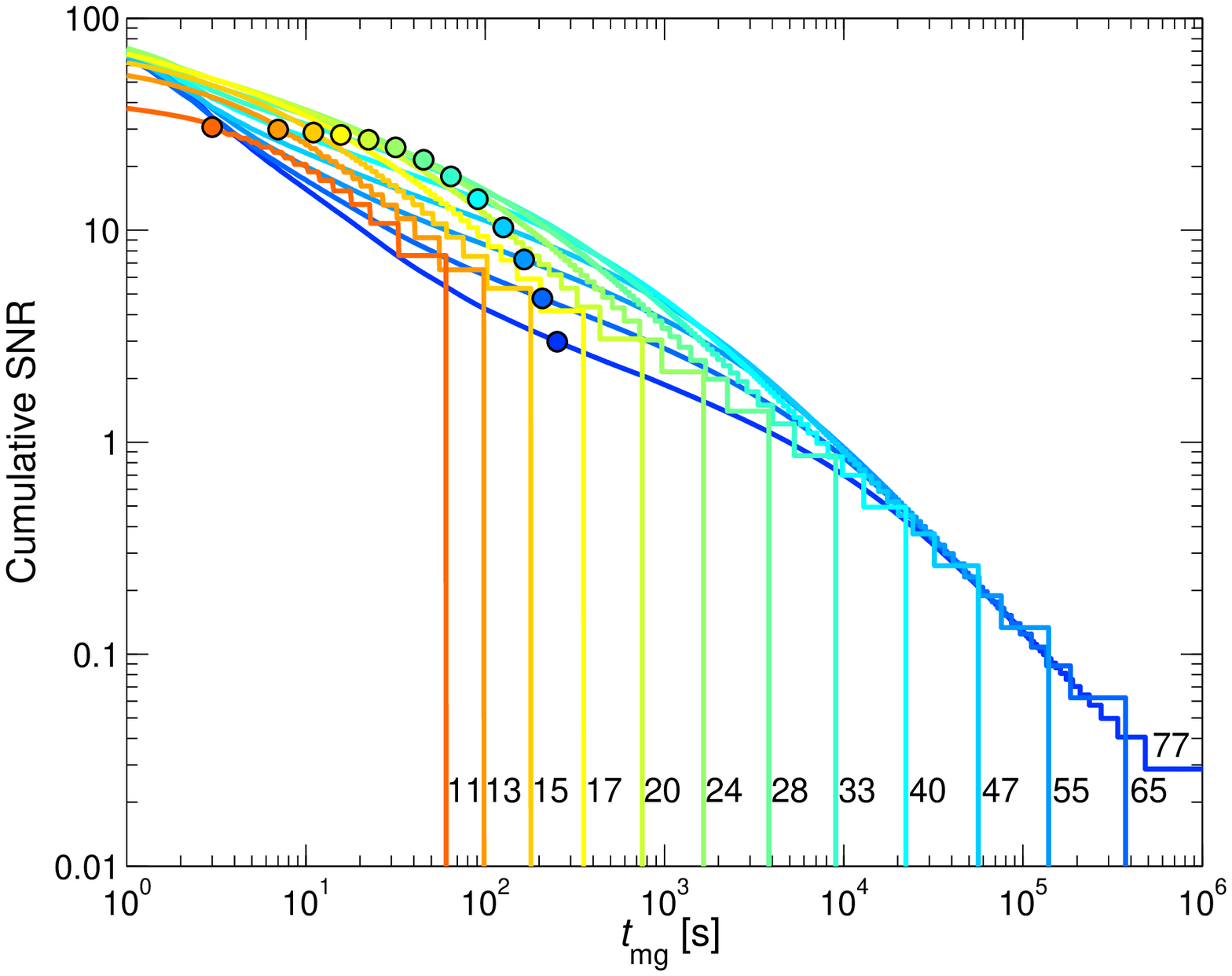}}
\mbox{
\includegraphics[width=8.5cm]{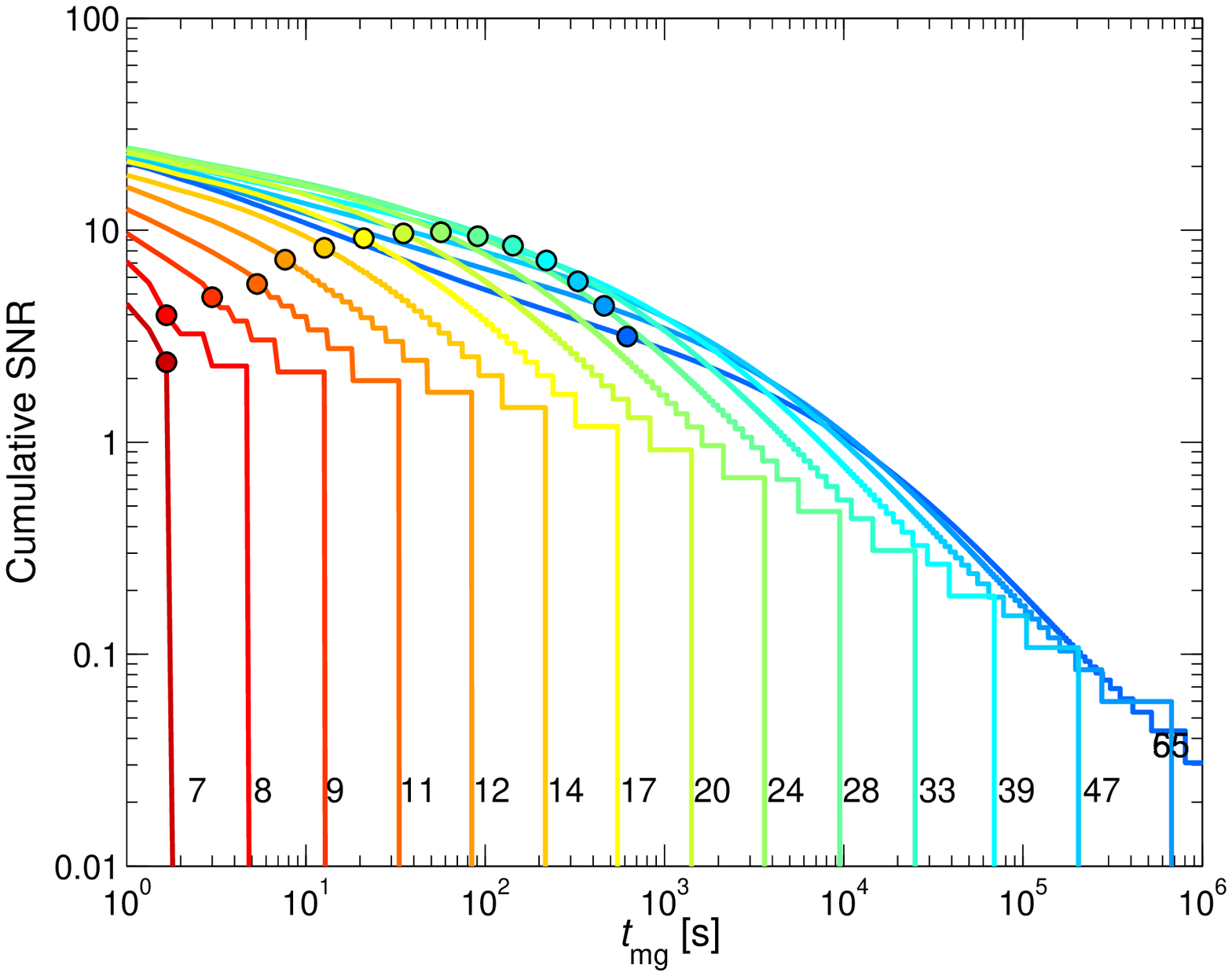}
\includegraphics[width=8.5cm]{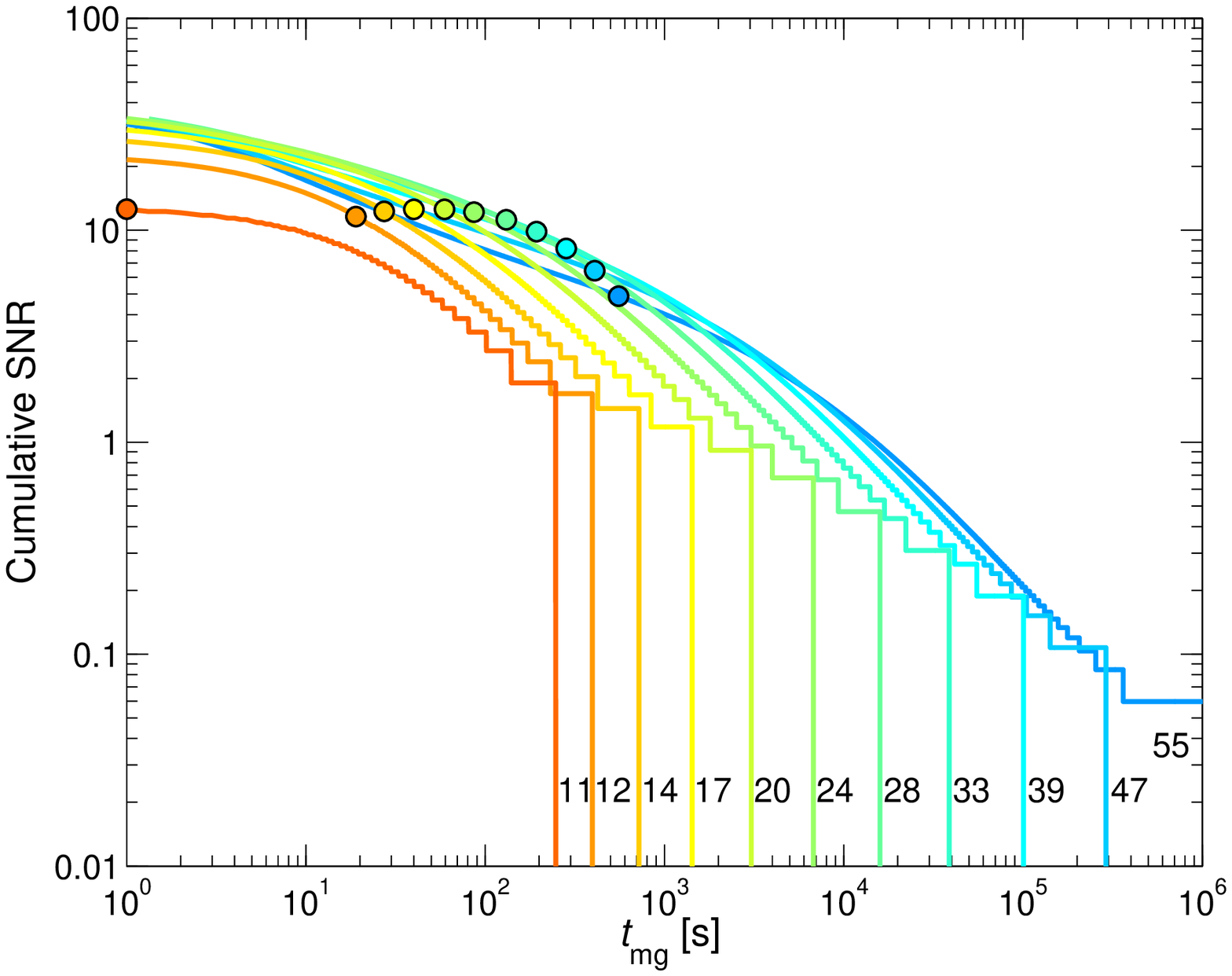}}
}
\caption{\label{f:snr-t}
Angular-averaged cumulative SNR for Advanced LIGO from the first passage as a function of time to merger
for 2.5PN (left) and 3.5PN (right) calculations and binary masses $m_1=m_2=10\Msun$ (top) and
$m_1=0.1m_2=1.4\Msun$ (bottom). Different curves correspond to different initial pericenter distances
as labelled, same as in Figs.~\ref{f:eccentricity-time} and \ref{f:orbits}. Far from the merger, the binary
is in the RB phase, and the SNR accumulates during close approaches. Circles resemble the
transition from the RB to a continuous chirp signal, where the orbital time is $0.5\,$s.
All panels assume a source at $100\,$Mpc, and average binary orientation.
}
\end{figure*}
Next, we split the GW signals into approximately 1 second segments (see Appendix \ref{app:FFT} for details)
and calculate the { angular-averaged} SNR for Advanced LIGO in each segment.
Figure~\ref{f:snr-t} shows how the SNR accumulates when measuring the signal from the first
passage to a time $t_{\rm mg}$ before merger for binaries at $100\,$Mpc.
Different curves correspond to binaries with different
pericenter distances at first passage $r_{p0}$ as labelled (same as in Figs.~\ref{f:eccentricity-time} and
\ref{f:orbits}). Different panels correspond to different binary masses and PN order (see figure caption).
Initially, in the RB phase of the binary evolution, the SNR accumulates in discrete
bursts. Although hard to see on the logarithmic scale, the strength of successive bursts is nearly equal.
Individual bursts are detectable separately at high significance only if the first passage is sufficiently close
($r_{p0}<15$ for a sky-averaged $S/N>5$ for $m_1=m_2=10\Msun$ at $100\,$Mpc).
The trends and order of magnitudes are broadly consistent with KGM.
The end point of the RB phase is marked with big circles, where the GW signal becomes continuous, starting
the final chirp.

The figure shows that the total SNR of the RB phase can be a substantial fraction of the
total SNR. This prediction is robust in both 2.5PN and 3.5PN calculations. {\it Therefore we conclude that a
coherent search for the train of bursts has a potential for detection even if the SNR of individual bursts
is small.}\footnote{Note that this implies that the RB phase does not show up clearly on time-frequency plots
of the SNR, since the signal power is small during each passage. Such
time-frequency plots are more useful for individual burst searches or for the final chirp,
when the total signal duration is not very large, and the SNR per unit time is substantial. }

\subsection{Total signal to noise ratio}
\begin{figure*}
\centering{
\mbox{
\includegraphics[width=8.5cm]{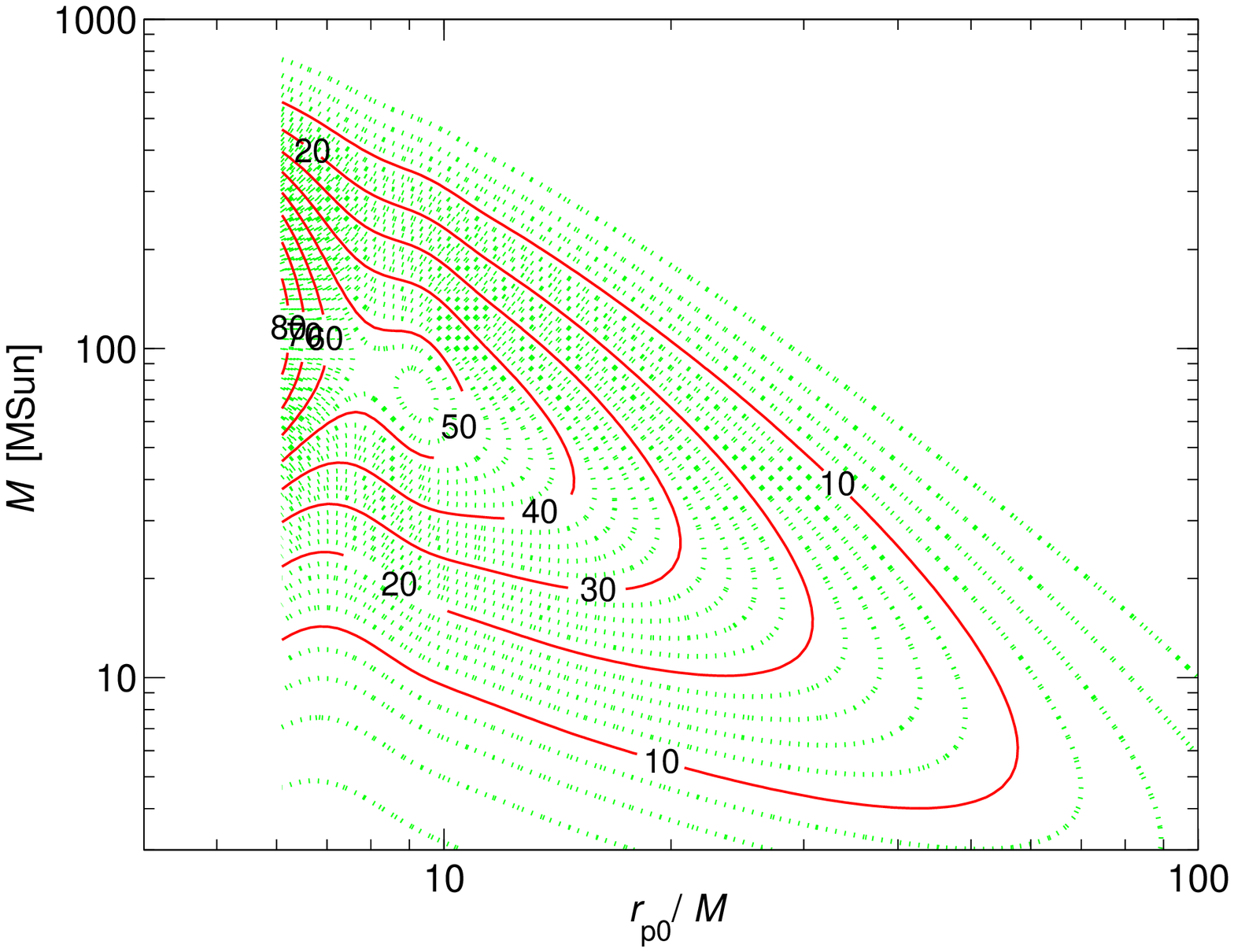}
\includegraphics[width=8.5cm]{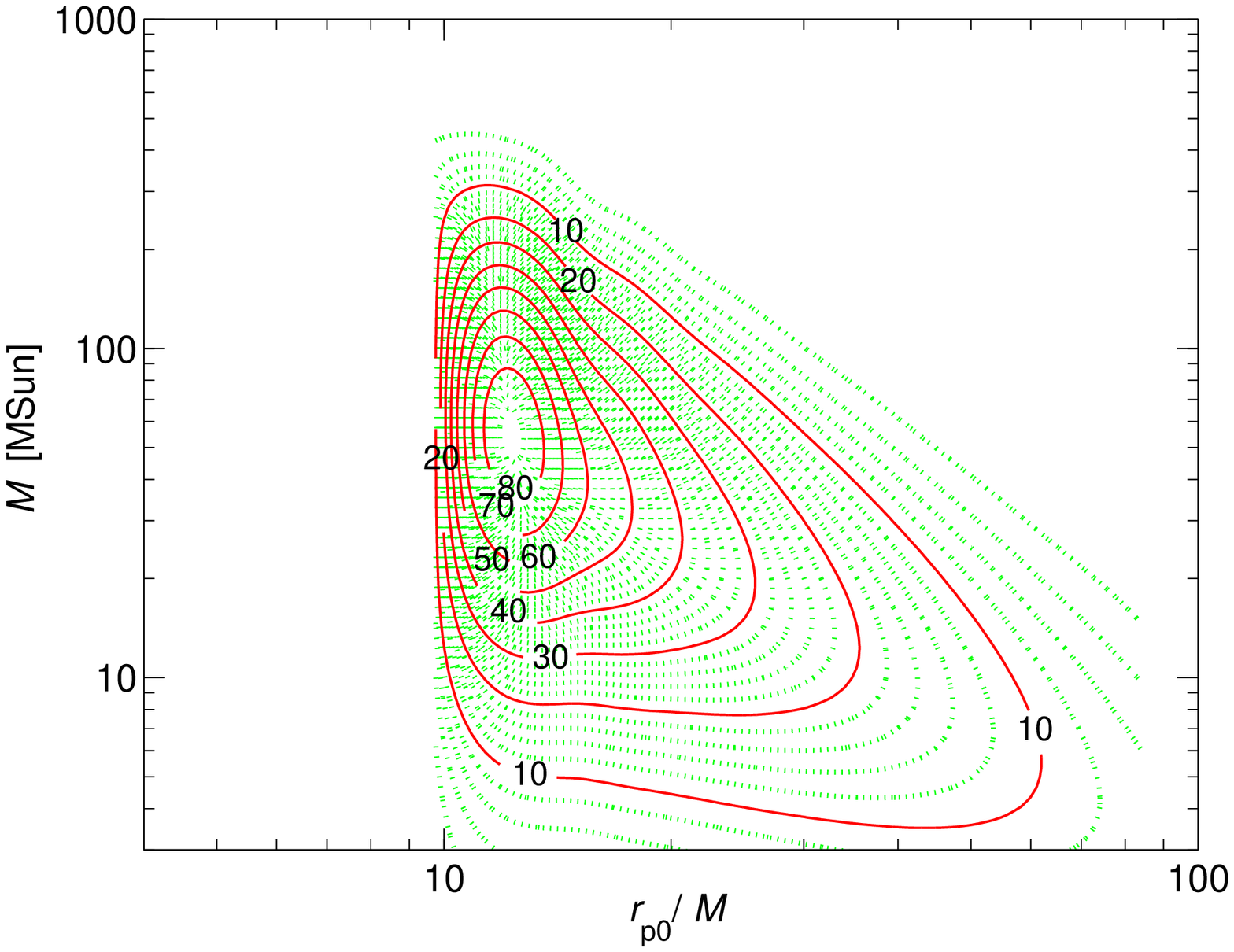}}
\mbox{
\includegraphics[width=8.5cm]{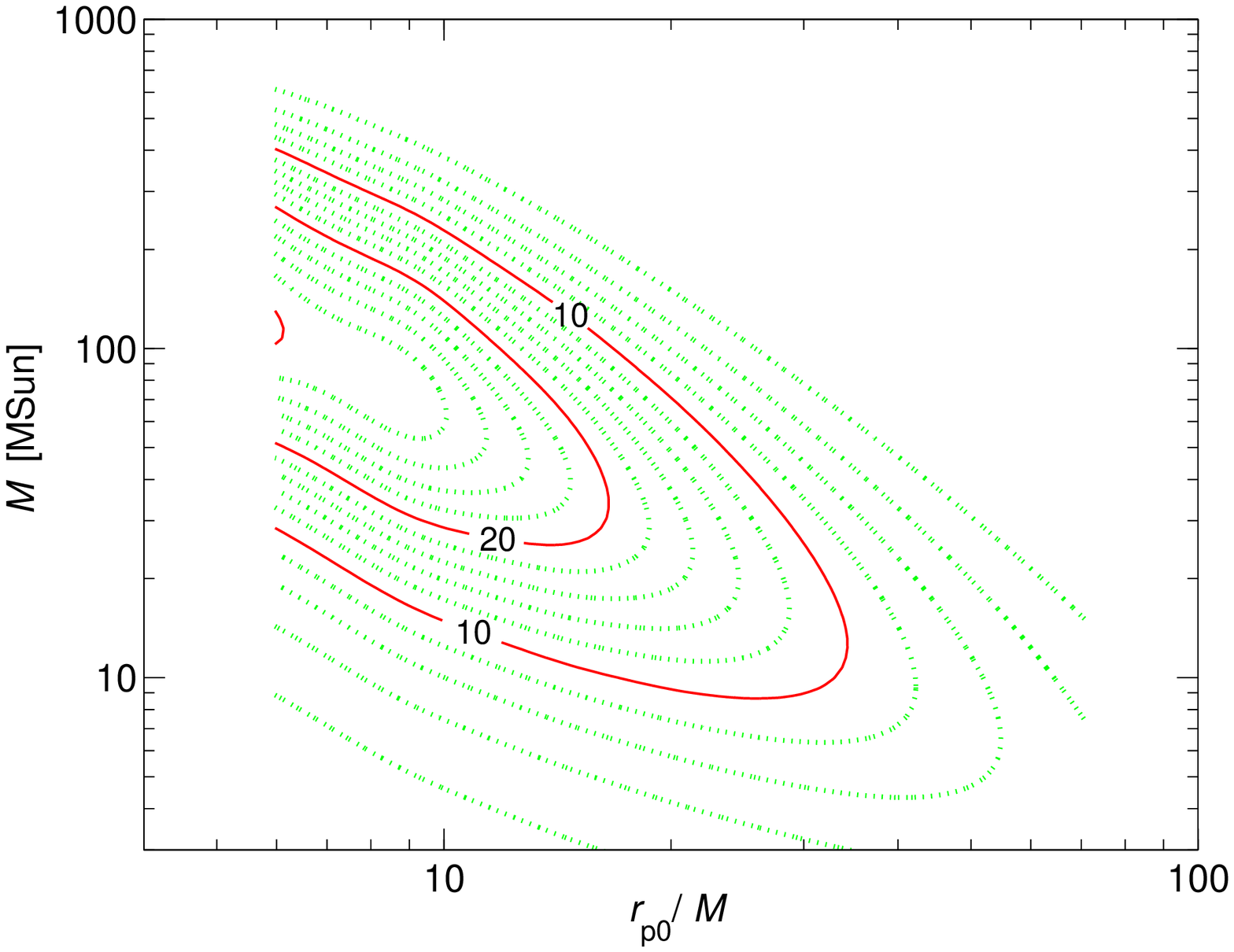}
\includegraphics[width=8.5cm]{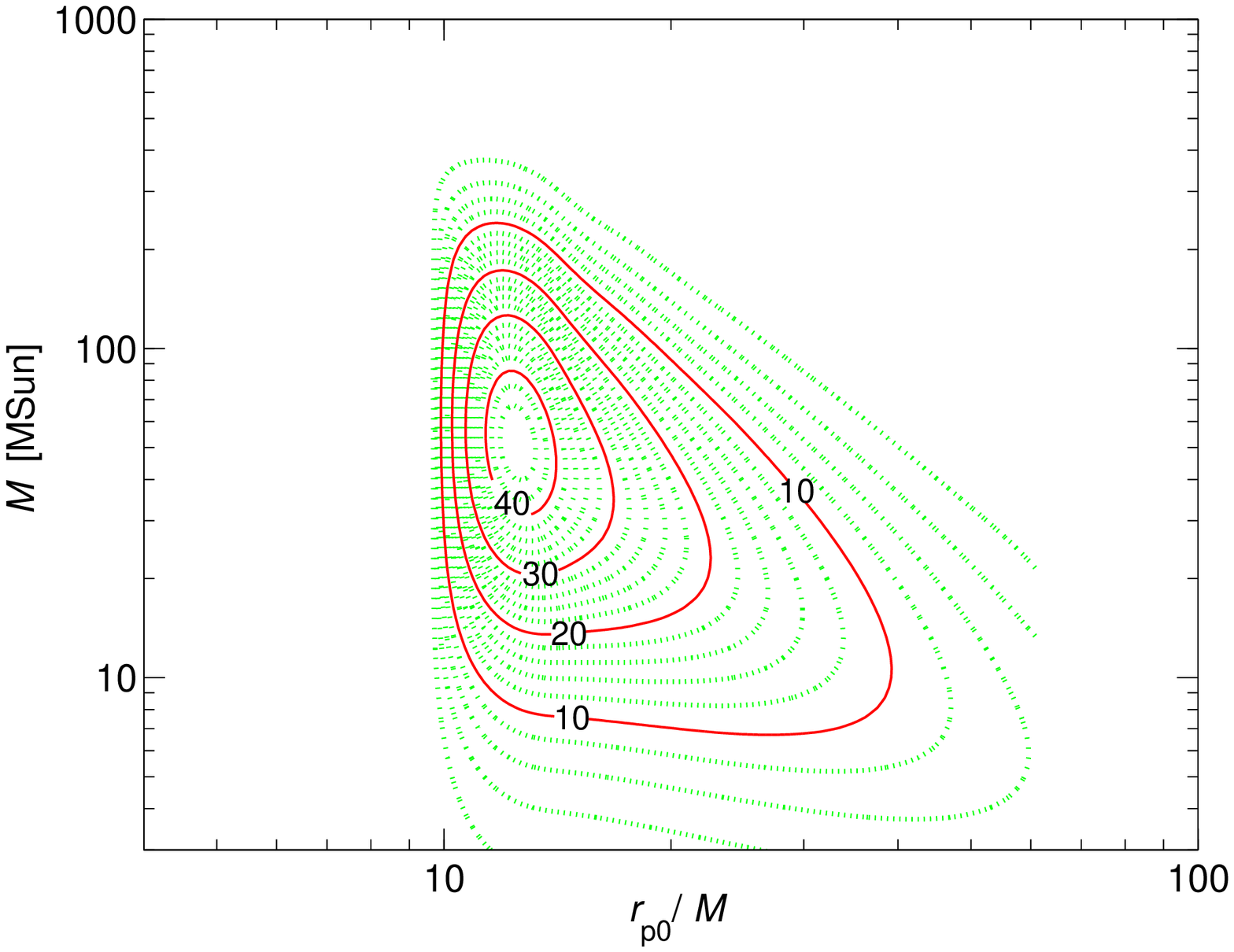}}
}
\caption{\label{f:snr}
SNR for Advanced LIGO in the RB phase when $e\gtrsim 0.6$,
as a function of the pericenter distance of the first approach $r_{p0}$ and binary mass
for mass ratio $q=1$ (left) and $0.1$ (right panels), in the 2.5PN (top panels) and 3.5PN calculation (bottom panels).
The source distance is $100\,$Mpc. For other distances the SNR is reduced proportionally.
}
\end{figure*}
\begin{figure*}
\centering{
\mbox{
\includegraphics[width=8.5cm]{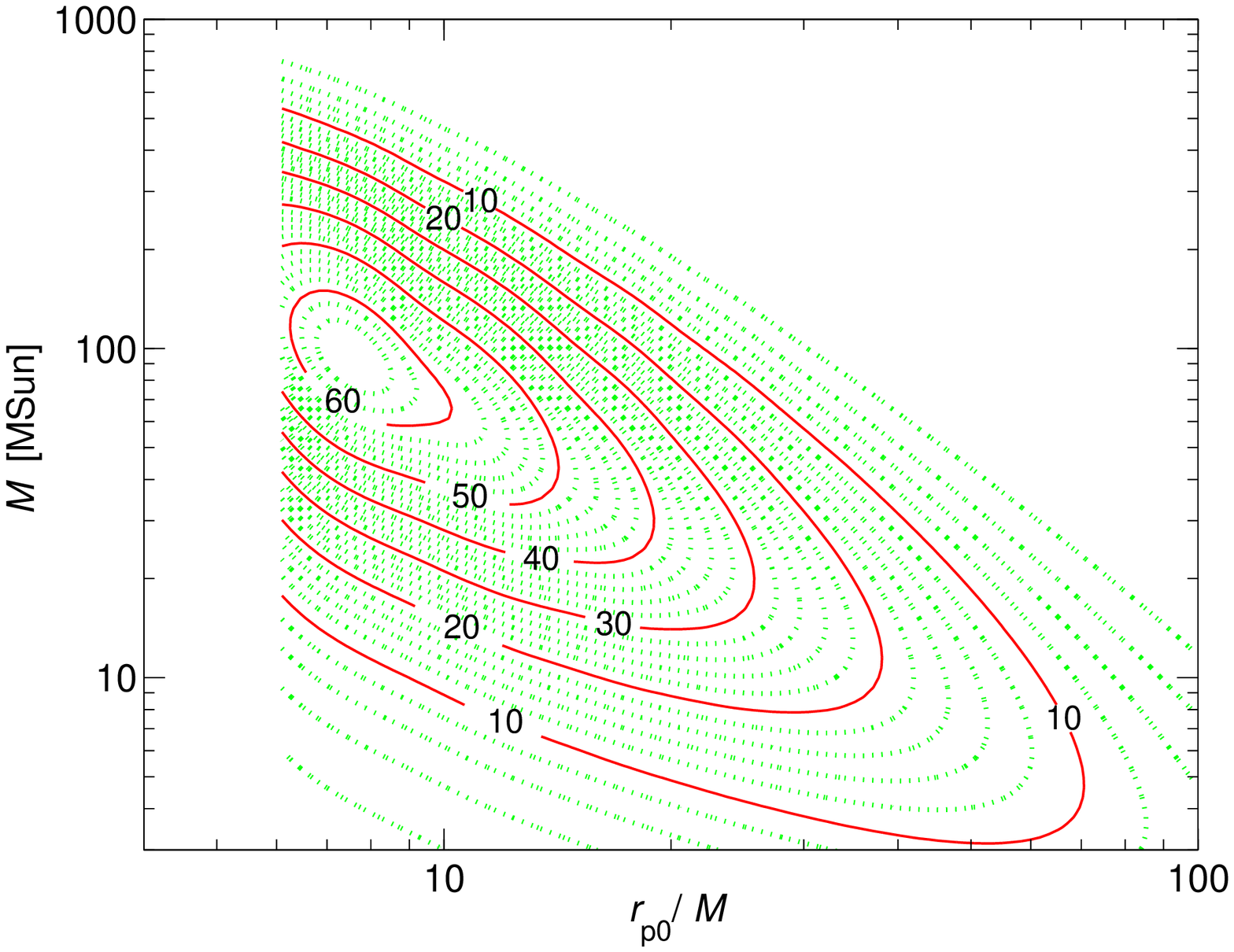}
\includegraphics[width=8.5cm]{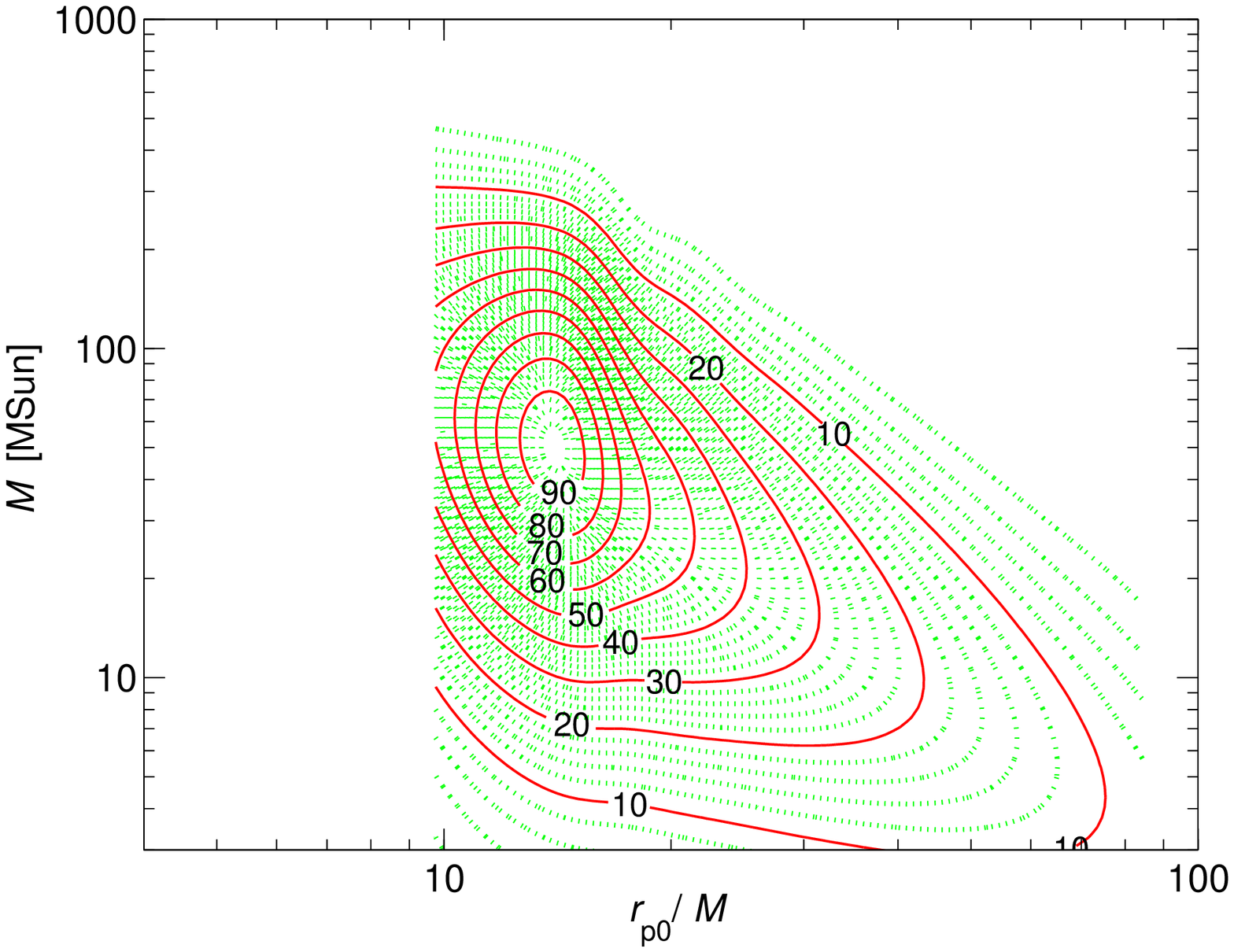}}
\mbox{
\includegraphics[width=8.5cm]{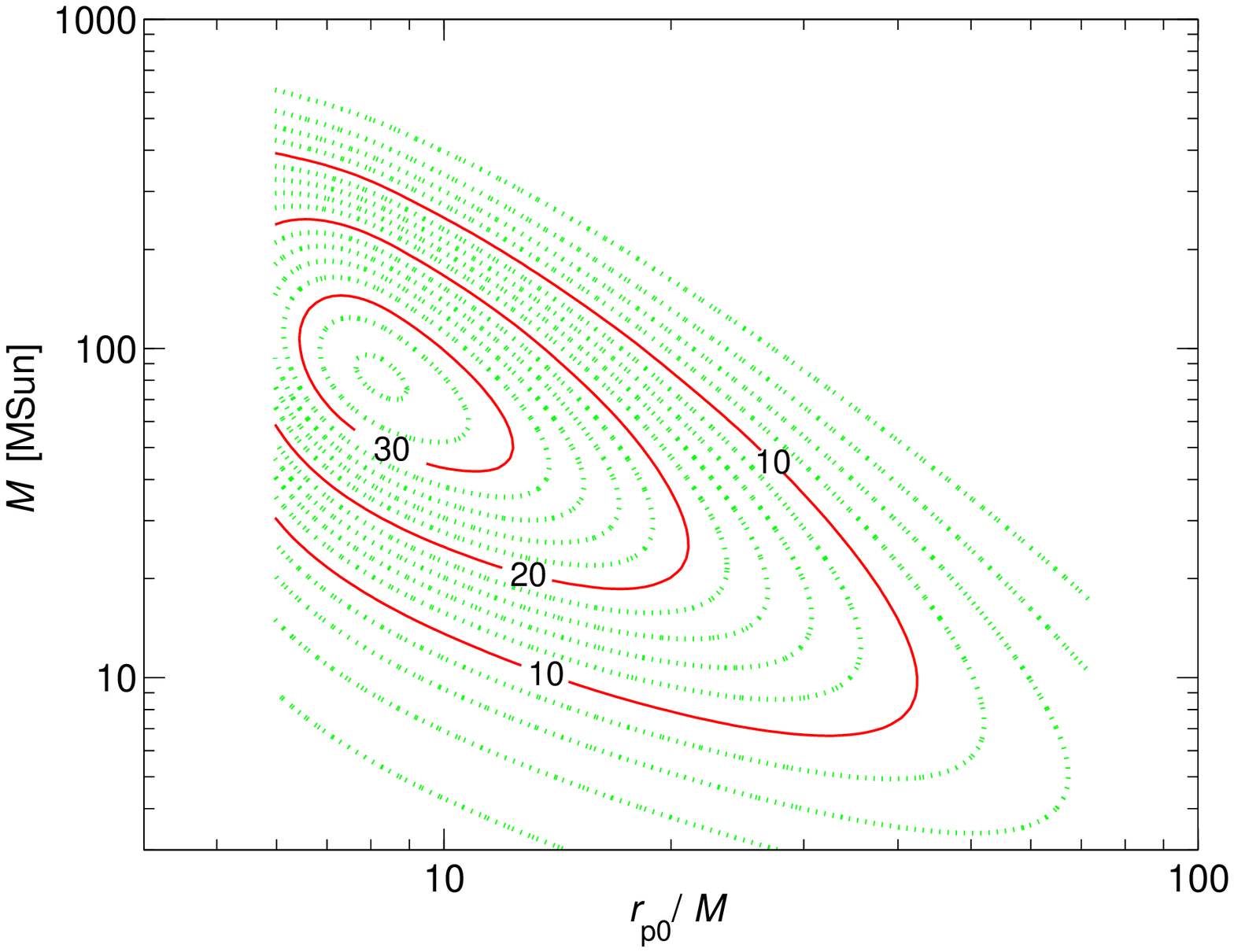}
\includegraphics[width=8.5cm]{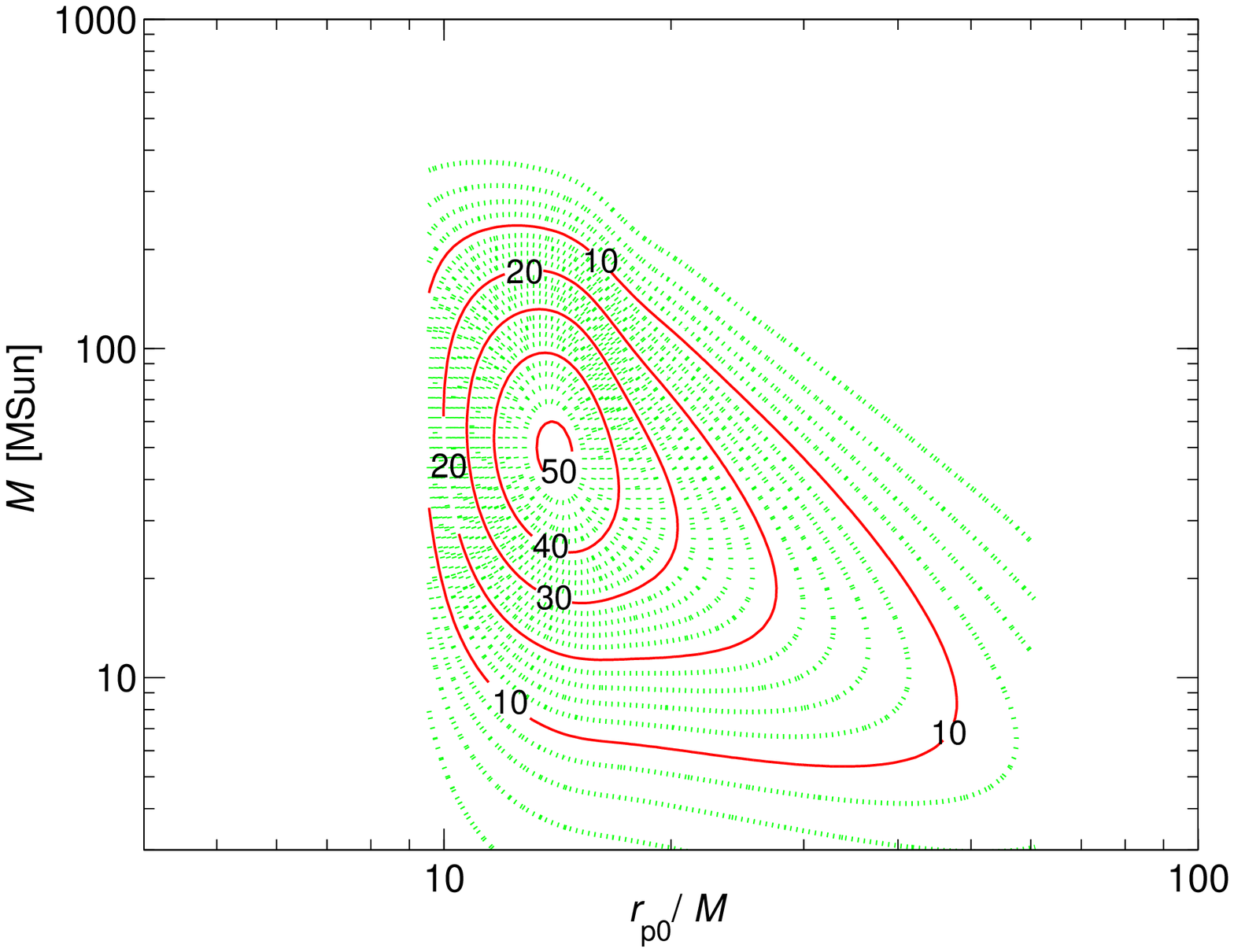}
}
}
\caption{\label{f:snr2}
Same as Fig.~\ref{f:snr} but for $e\gtrsim 0.45$, corresponding to $f_{p}/f_{\rm orb}\lesssim 1/3$.
}
\end{figure*}

Up to this point we restricted to a single choice of masses for BH/BH and BH/NS mergers. Now we
 extend the
analysis to all possible total masses observable to LIGO between $2\Msun$ to $1000\Msun$.

The contours in Figures~\ref{f:snr} and \ref{f:snr2} show the angular-averaged SNR for Advanced LIGO in the RB phase
for sources at $100\,$Mpc, as a function of $r_{p0}$ and $M$, for mass ratio $q=1$ and $0.1$ (top and bottom panels),
for 2.5PN and 3.5PN calculations (left and right panels). In Figures~\ref{f:snr} and \ref{f:snr2}, we select the segment of the signal
where $e\gtrsim 0.6$ and the $e\gtrsim 0.45$, respectively, and evaluate the corresponding numerical FFT and SNR
for Advanced LIGO. Note that the pericenter passage timescales
in these cases are at least 5 times and 3 times shorter than the orbital timescale, so the GW signal consists of well separated
RBs between silent periods.\footnote{Note however, that the time duration between individual bursts may be smaller than
0.5 sec in this case, so the circles in \ref{f:eccentricity-time}, \ref{f:orbits}, and \ref{f:snr-t} do not coincide with
$e=0.6$ or $e=0.45$, see Fig.~\ref{f:orbits}.}
The SNR results for the 2.5PN and 3.5PN calculations in the RB phase are consistent to within $30\%$.
The 3.5PN curves cannot resolve the SNR when $r_{p0}\lesssim10$ for reasons already
mentioned (namely, the poorly behaved approximation).
We expect the SNR to be significantly larger for zoom-whirl orbits with $r_{p0}\lesssim 10 M$ relative to our calculations
(KGL). As shown by Fig.~\ref{f:AdLIGO-spectra-circ}, the 2.5PN calculation systematically
underpredicts the SNR at all frequencies by a factor $\sim1$--2 for circular orbits at frequencies  $f\lesssim 0.3\, f_{\rm ISCO}$,
while it may be even more uncertain at higher frequenciesc closer to ISCO.

Comparing Fig.~\ref{f:snr} to Fig.~11 of OKL, we see that a considerable fraction of the total SNR is in the RB phase,
for masses less than $\sim 20\Msun+20\Msun$ and initial pericenter distances $r_{p0}\lesssim 40$, making the
RB signals typically detectable to several $100\,$Mpc with high significance. Our highest SNR results
in the RB phase, are typically a factor 2--3 lower than OKL's for the full waveforms.
While OKL claimed that intermediate mass BHs with $m_1=m_2=400\Msun$ are
detectable to $1\,$Gpc with ${\rm SNR}=5$ with Advanced LIGO for $r_{p0}\sim 6 M$, our 2.5PN and 3.5PN calculations cannot
accurately model orbits in this range to either confirm or rule out such claims.

\subsection{Precession effects}

Figures~\ref{f:AdLIGO-spectra-ecc1}--\ref{f:snr-t} assume orientation and polarization averaged waveforms
for zero spin. In reality, however, each GW detector will be sensitive to a single linear combination of the
$+$ and $\times$ polarizations.
The measured GW signals are strongly modulated by  the 1PN GR precession of the eccentric orbit
within the orbital plane. Further, if the objects are spinning with general non-aligned
spin orientations, spin-orbit precession further modulates the signal waveform  at 1.5PN order \citep{1995PhRvD..52..821K}.
In the leading order flux averaged approximations, the precession periods are respectively,
\begin{align}
 t_{\phi} &= \frac{2\pi }{3} \frac{r_{p}^{5/2}}{M^{3/2}} \frac{1+e}{(1-e)^{3/2}} \nonumber\\&
= 1.0\,{\rm s} \times  \frac{M}{20\Msun} \left(\frac{r_p}{30   M}\right)^{5/2} \frac{1+e}{(1-e)^{3/2}}\,, \nonumber \\
 t_{\Psi} &= \pi \frac{r_{p}^{3}}{M^2} \left(\frac{1+e}{1-e}\right)^{3/2}  \nonumber\\&
= 8.4\,{\rm s} \times  \frac{M}{20\Msun} \left(\frac{r_p}{30 M}\right)^{3}\left(\frac{1+e}{1-e}\right)^{3/2}
\,.\end{align}
{ In contrast, the total signal duration after the first flyby is approximately (see Eqs.~(13) and (27) in OKL)
\begin{align}
 t_{\rm mg} &= \frac{M}{\sqrt{4\pi}} \left(\frac{3}{85\eta} \right)^{3/2} \left(\frac{2 r_{p 0}}{M}\right)^{21/4}\nonumber\\
&= 0.89\,{\rm hr}\times (4\eta)^{-3/2} \frac{M}{20\Msun} \left(\frac{r_{p 0}}{30 M}\right)^{21/4}\,.
\end{align}
The Earth spin also modulates the measured waveform, which could be significant for
waveforms lasting several hours, i.e. if $r_{p0}\gtrsim 30 M \times (M/20\Msun)^{-4/21} (4\eta)^{2/7}$.}
Clearly, the evolution time in the RB phase is almost always much larger than the precession timescales
$t_{\phi}$ and $t_{\Psi}$ (see Fig.~\ref{f:snr-t}). Accounting for these effects is going to be crucial for detection algorithms.

\subsection{Coincident multiwavelength observations}
\label{s:multiwavelength}
\begin{figure}
\centering{
\mbox{\includegraphics[width=9cm]{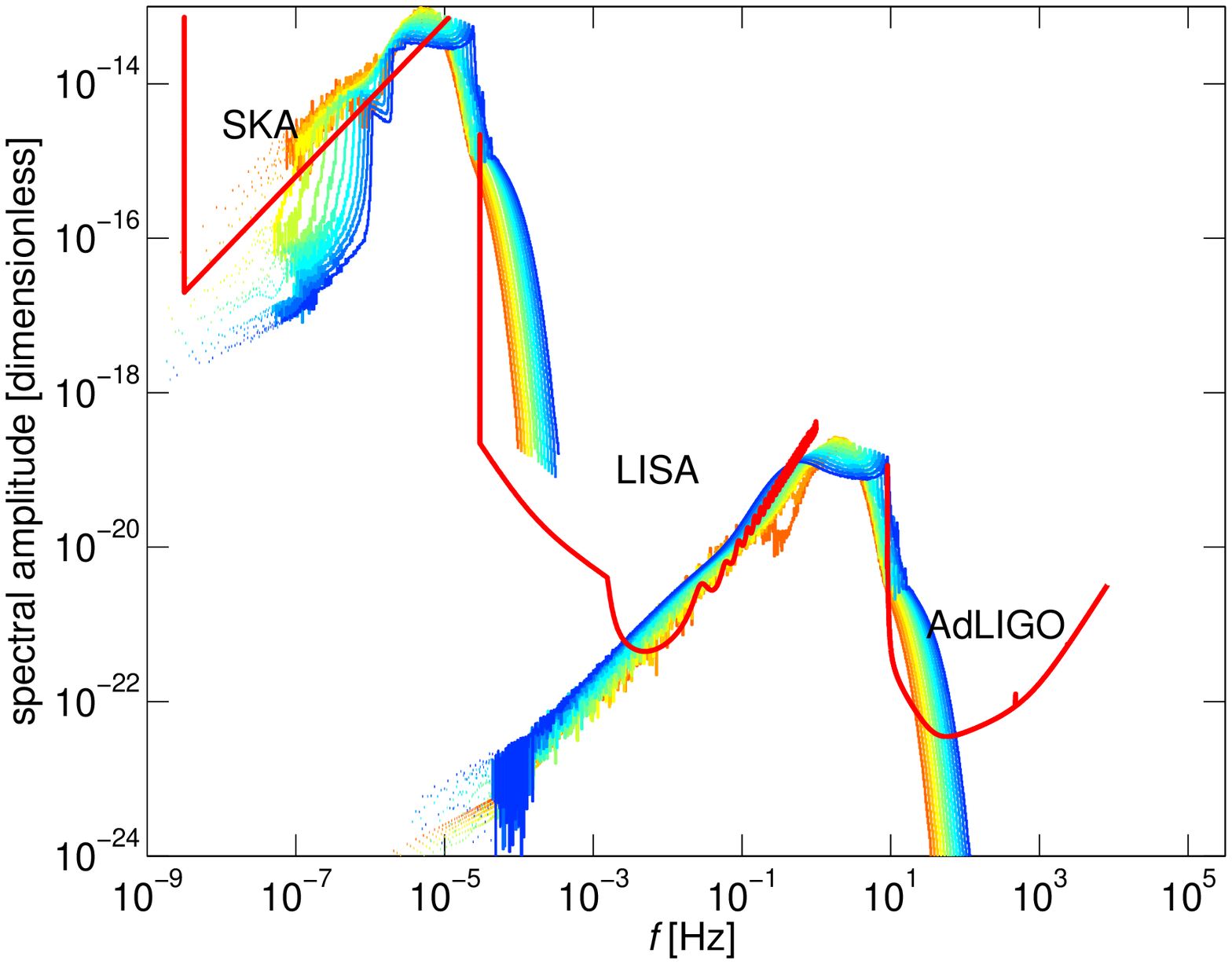}}}
\caption{\label{f:AdLIGO-LISA-PTA}
The GW spectra for supermassive and intermediate-mass BH binaries. The signal is marginally in range for coincident
multiwavelength GW detections with LIGO, LISA, and SKA. Different overlapping curves
represent different impact parameters corresponding to Figs.~\ref{f:orbits} and \ref{f:eccentricity-time}.
{ The maximum observation time is limited to 10 years.}
}
\end{figure}
An interesting unique characteristic of these GW signals is that they are broadband, and can be
detected with multiple GW detectors coincidentally in different frequency bands.
Figure~\ref{f:AdLIGO-LISA-PTA} shows the GW spectra for an optimally oriented (i.e. face-on) binary for
total masses $200\Msun \lesssim M \lesssim 720\Msun$ (intermediate mass black holes IMBH)
and $7\times10^7\Msun \lesssim M \lesssim 3\times 10^8\Msun$ (supermassive black holes SMBH)
with different impact parameters  at 100 Mpc. { For SMBHs, the maximum observation time is limited
to the final 10 years  before coalescence.} The figure shows that the spectral range of the signal
spans the frequency range of multiple instruments for these masses for a wide range of initial
$r_{p0}$. Ultimately, the frequency range of the Einstein Telescope is most ideal to detect these
coincident LIGO/LISA sources with IMBHs. Going to even higher masses and lower frequencies,
we find that parabolic GW captures of SMBHs are detectable with future pulsar timing arrays (PTAs)
such as the SKA and LISA, coincidentally.

The event rates of such IMBH or SMBH encounters within a few tens of Mpc is currently unknown.
Portegies-Zwart et al. \cite{2006ApJ...641..319P} predict that the galactic centers hosts 50 IMBHs,
which could undergo GW capture events. GW observations of these events could prove the existence of
such a population.
Regarding SMBHs, they are much more luminous and easier to identify using electromagnetic (EM) observation.
There is an ongoing effort to search for SMBH binaries using EM observations.
3C 66B, a nearby radio galaxy at 80 Mpc, was interpreted as an SMBH binary with a total mass
$5.4\times10^{10} \Msun$ \cite{2003Sci...300.1263S}. However, Jenet et al.~\cite{2004ApJ...606..799J}
have shown that the GWs would be detectable with PTAs but are not observed, ruling out the
SMBH binary interpretation for this particular source. Based on our SNR estimates, we conclude that
PTAs could search for RB sources at similar distances.

The prospects for multiwavelength observations may be somewhat better
in reality than shown in Fig.~\ref{f:AdLIGO-LISA-PTA}. Our calculations are artifically truncated at relatively small frequencies
$f<0.4f_{\rm ISCO}$ due to the inaccuracy of the simulations at small seperations $r\lesssim 10M$.
The complete signal, including the final chirp and the following ringdown, extends to higher
frequencies than shown in Fig.~\ref{f:AdLIGO-LISA-PTA}, and extending into the LIGO and LISA bands
for IMBH and SMBH binaries, respectively.
As the inspiral and ringdown phases are both detectable,
these sources constitute a new class of ``golden binaries'', and may be useful to probe
strong field gravity \cite{2005ApJ...623..689H}

\section{Discussion}
We have examined the evolution and GW spectra of eccentric binaries formed by GW emission using the
2.5PN and 3.5PN equations of motion of Will and collaborators \cite{2002PhRvD..65j4008P,2004PhRvD..69j4021M,2005PhRvD..71l9901M,2005PhRvD..71h4027W,2007PhRvD..75f4017W,2011CQGra..28q5001L}.
The capture cross section, evolutionary tracks, and the GW spectra are in remarkable agreement with the
simple analytic estimates of OKL. The 2.5PN and 3.5PN results bracket those in OKL.

After the formation of the binary, the GW signal is described by a long repeated burst (RB) phase
lasting minutes to days, followed by a continuous powerful chirp. The signal evolves from the RB to the chirp phase
within the frequency band of Advanced-LIGO type instruments. The maximum distance of detection for waveforms
in the RB phase is around 300--600 Mpc for Advanced LIGO for an average orientation BH/BH binary
with ${\rm SNR}\sim 5$--10 for $10\, M \lesssim r_{p0}\lesssim 25\, M$ and
$20\Msun\lesssim M\lesssim100\Msun$ (see Figs.~\ref{f:snr} and \ref{f:snr2}).
We find that the total SNR is substantial already in the RB phase
when the eccentricity and separation are relatively large. Numerical relativity may best resolve the prospects
for detecting the GW signal in the final powerful chirp phase, or signals with smaller impact parameters
leading to zoom-whirl orbits.

We found that relativistic corrections do not greatly modify the event rate estimates of these waveforms.
OKL have shown that the Advanced LIGO detection rates for these sources may be around $1$--$3000\,{\rm yr}^{-1}$, depending on the
number of stellar mass BHs with masses between $10$--$40\Msun$ in galactic nuclei.
The event rates for IMBH encounters and encounters involving a BH and NS may be equally numerous.
Discarding the final powerful chirp, and assuming conservatively that sources are detectable in the RB phase
to 1/3rd of the distance of the full signal, the event rates for these sources may be reduced by a factor $27$,
to give a total detection rate ${\cal R}_{\rm RB} \sim 0.03$ -- $100\,{\rm yr}^{-1}$.
The event rates in reality may be much higher than these estimates which in OKL was based on an isotropic density distribution
of compact objects in galactic nuclei. However, heavier objects segregate into very anisotropic configurations
efficiently through vector resonant relaxation  \citep{1996NewA....1..149R}. Indeed, the observed distribution
of massive stars in the Galactic center is anisotropic, comprising two disks with a thickness of
$10^{\circ}$ \citep{2009ApJ...697.1741B}. This thicness is consistent with the prediction of mass segregation
for objects with these masses in statistical equilibrium with lighter $1\Msun$ stars. It is plausible to expect
a similar anisotropic distribution of BHs in galactic nuclei. Since the event rates are proportional to the squared density
of compact objects, one might expect that the true event rates are larger by up to a factor 100 than in OKL, making
these sources much more numerous than other LIGO-VIRGO sources \cite{2010CQGra..27q3001A} (Kocsis \& Tremaine in preparation).

We have shown that encounters between IMBHs may be detected with Advanced LIGO and LISA coincidentally
if the source is within $50\,$Mpc. Similary for SMBHs, coincident detections may be possible with LISA and
future Pulsar Timing Arrays such as the SKA.

We have also estimated the SNR for detecting the GWs from BH/NS encounters, and found that  a detection with
${\rm SNR}=10$ may be possible to $300\,$Mpc. These encounters may lead to tidal
disruption events and may exhibit luminous coincident electromagnetic variations \cite{2010ApJ...720..953L}.
Indeed, short-hard gamma ray bursts (GRBs) are
modelled as the merger of two NSs or a NS into a stellar mass BH, following a circular inspiral.
It is plausible to expect that GW captures
leading to eccentric coalescences also result in similar phenomena. An important difference for very eccentric orbits
is that the pericenter separation may be smaller than the ISCO for circular orbits \cite{2011arXiv1105.3175S}.
This may lead to tidal stripping, { partial disruption,  or the shattering of the NS crust during close approaches
\cite{2005MNRAS.356...54D,2012PhRvL.108a1102T}}.
As the mass of the NS is reduced, its radial size increases, so
that tidal stripping becomes more efficient for successive close approaches.
GRBs with observed precursors \cite{2010ApJ...723.1711T} might correspond to these eccentric events.
Thus, this process has a potential to generate electromagnetic bursts tracking the GW signal.
The LIGO data near GRBs could be searched for these particular GW signals.
If such counterparts are successfully identified, such processes could be used as standard sirens to
constrain the cosmological model \cite{2006PhRvD..74f3006D,2006ApJ...637...27K}, and the mass of the graviton
in alternative theories of gravity \cite{2008ApJ...684..870K}.

Our 2.5PN and 3.5PN calculations demonstrate the slow convergence of the PN expansion for these encounters
in the final inspiral phase.
More accurate calculations would be necessary to make more accurate predictions on the detectabiltiy of these
signals. Indeed, the waveform modelling precision may be an important limiting factor for concrete detection techniques
and parameter measurement accuracy. The theoretical errors due to the imperfect modelling of these signals may be dramatic
for sources with small initial pericenter distances or during the later parts of the signals approaching merger \cite{2007PhRvD..76j4018C}.
However, we have shown that the initial RB phase of the GW signal carries a considerable total SNR.
While the theoretical modeling of the GW signal in this phase may be more accurate, their detections requires
searching for a train of GW bursts over long timescales which have individually a small amplitude. The
standard LIGO-Virgo detection pipeline is not sensitive to these signals. A long-duration transient search with
a network of instruments might be a more promising avenue for detection \cite{2011PhRvD..83h3004T}.
However, as the expected waveforms are well described in the RB phase, optimized data analysis
techniques could be developed for their targeted detection.

Future studies should investigate how accurately can a network of
detectors measure the physical parameters of these sources.
These are the component masses, binary distance from the Earth, binary orientation, sky location, time and phase at merger,
initial pericenter distance (or equivalently, the impact parameter), and the initial velocity before the first encounter. The later two
parameters $(r_{p0},w)$ affect the eccentricity evolution, which is nonnegligible when the signal is in the detector's frequency band.
Measuring the eccentricity evolution yields $r_{p0}$, and is mostly insensitive to $w$.
The later might be hard to detect directly, unless the GW signals from the first few passages can be resolved.
However, since binary capture implies a maximum $r_{p0}$ for a fixed $w$, an estimate of
$r_{p0}$ puts an upper bound on $w$. This may already be sufficient to distinguish between
sources in galactic nuclei where typically $v\gtrsim 1000\,{\rm km/s}$ from those in globular clusters where
$v\lesssim 60\,{\rm km/s}$ (OKL). Regarding sky localization, the modulation
related to GR and spin precession or the rotation of Earth,
can be a substantial help, as the GW signal amplitude changes significantly due to these effects
during these long duration signals.
We expect this to greatly improve the measurement accuracy for these eccentric sources beyond the
$10\,{\rm deg}^2$ accuracy of regular circular inspirals \cite{2011arXiv1105.3184N}.

\acknowledgments
We gratefully acknowledge valuable discussions with Szabolcs Marka who suggested this collaboration, Sean McWilliams,
Scott Hughes, Alessandra Buonanno, and Ryan O'Leary.
This work was supported by an NSF grant AST-0908365.
JL gratefully acknowledges
support of a KITP Scholarship,
under Grant no. NSF PHY05-51164.
BK acknowledges support from NASA through Einstein Postdoctoral
Fellowship Award Number PF9-00063 issued by the Chandra
X-ray Observatory Center, which is operated by the Smithsonian
Astrophysical Observatory for and on behalf of the National
Aeronautics Space Administration under contract NAS8-03060,
and the Hungarian Research Fund  OTKA (grant 68228).

\appendix
\section{Angular average waveforms}
\label{app:angular}

We calculate the numerical waveforms in a direction along the orbital axis of the binary, and use it for a proxy for general inclinations.
In particular, in the circular orbit-averaged case, the radiation is dominated by the $(\ell, m)=(2,2)$ spin-2 weighted spherical harmonic,
so that $h_{+}(t)=(1+\cos^2\iota) A \sin(2\pi f t)$ and $h_{\times}(t) =  (2\cos \iota) A \cos(2\pi f t)$
as a function of the inclination angle relative to the line of sight. Here $A$ gives the amplitude scale, which
in the optimal orientation $\cos \iota =1 $ satisfies $8 A^2 \equiv (h_{+}^2 + h_{\times}^2)_{\rm opt} $.
Averaging over an isotropic distribution, $2 \langle h_{+}^2\rangle = (28/15) A^2$ and
$2 \langle h_{\times}^2\rangle = (4/3) A^2$, so that $\langle h_+^2 + h_{\times}^2 \rangle = (2/5) (h_+^2 + h_{\times}^2)_{\rm opt}$,
the GW power in the optimal orientation is 2.5 times larger than the average GW power.
We find that these identities are satisfied to within $1\%$ accuracy
in the Peters-Matthews leading order orbit averaged flux calculation for arbitrary excentricity. We will use them
to change from the calculation evaluated at the optimal orientation to estimate the GW signal in the ``average'' case.

The angular-average waveform amplitude is then
\begin{equation}\label{e:htilde}
\tilde{h}^2(f) = \langle F_{+}^2 \tilde{h}_+^2(f)  + F_{\times}^2 \tilde{h}_{\times}^2(f)\rangle =
\frac{2}{25} [\tilde{h}_{+}^2(f)  + \tilde{h}_{\times}^2(f)]_{\rm opt}
\end{equation}
where we have used $\langle F_{+}^2 \rangle=\langle F_{\times}^2 \rangle= 1/5$
for the beam patterns and a factor $2/5$ for the inclination-averaged GW amplitude. While
in general, $\langle F_{+}^2 \tilde h_+^2 \rangle \neq \langle F_{+}^2 \rangle  \langle  h_+^2 \rangle $,
however, the error introduced by this approximation is only $6\%$ in the circular case, so we shall
use Eq.~(\ref{e:htilde}) as an approximation.

Note, that the signal measured by a single detector in the optimal orientation with $F_+=1$ and $F_{\times}=0$
is larger than $\tilde h(f)$ by
\begin{equation}
 \tilde h_{+,\rm opt}(f) = \frac{5}{2}\times \frac{1}{\sqrt{2}}\times\, \tilde{h}(f) = 1.8\, \tilde{h}(f)\,.
\end{equation}

In the stationary phase approximation, the angular averaged spectral amplitude is
\begin{align}\label{e:SPA}
 \tilde{h}(f) &= \frac{4}{5}\times\left(\frac{5}{96}\right)^{1/2} \pi^{-2/3} \frac{M^{5/6}\eta^{1/2}}{D} f^{-7/6}
\end{align}
assuming circular orbits, { leading PN order} orbit-averaged flux, and no spins \citep{1995PhRvD..52..848P}.
Here, the $4/5$ factor is the RMS of the polarization amplitude ${\cal A}_{p}=(1+\cos^2 \iota)F_{+} - 2i\cos\iota F_{\times}$
for random orientations. A similar approximate formula is available for eccentric orbits
summing over the contributions of orbital harmonics (OKL).

\section{Subtleties with numerical FFT}
\label{app:FFT}
There are some numerical  subtleties we have to consider when calculating the FFT of a very long data stream, of length $T$,
spanning several months in some cases with several kHz sampling frequency. To save computation time, we split the waveform into segments of size $\Delta T$.
side effects: (i) the spectrum becomes inaccessible at $f<2/\Delta T$, (ii) the frequency resolution will be discrete with a fundamental
frequency $1/\Delta T$, and (iii) it can introduce artificial features and numerical errors if the signal and its derivatives
do not vanish at the boundaries. To minimize these errors but optimize the computation time, we apply the following procedure:
\begin{enumerate}
 \item Keep the segment size as large as possible. In practice
$\Delta T_i\sim \Delta T \equiv 2\times 10^4 {\,\rm sec} (M/20 \Msun)$  typically for the $i^{\rm th}$ segment,
except for Fig.~\ref{f:snr-t}, where $\Delta T_i\sim \Delta T \equiv 1\,{\rm sec}$.
  \item Choose split points measuring them from the end of the timeseries (i.e. near merger),
so that the last split interval is at least size $\Delta T$.
  \item Adjust the split points between intervals to the nearest local minima of $h_{+}^2(t) + h_{\times}^2(t)$.
  \item Apply a gradual fade in and fade out near the edge of the waveform over a timescale $t_{\rm fade}\sim 500 M$.
  \item Resample the $h_+(t)$ and $h_{\times}(t)$ time series with uniform time steps.
This is necessary since the simulations use adaptive time steps.
  \item If the orbital time $t_{\rm orb}$ is larger than $\Delta T$, then truncate the data stream to the pericenter passage,
centering  the split window of duration $T$ there.
  \item Append the data stream with $h_{+}(t)=h_{\times}(t)=0$, if the total signal duration $T$ is smaller than $\Delta T$.
 \item Calculate the FFT of the resulting time series, corresponding to $h_+(t)$ and $h_{\times}(t)$, for each time segment.
  \end{enumerate}

{
\section{Event rates}\label{app:rates}
Here we provide a simple estimate of the event rates of these sources in galactic nuclei, and highlight the main sources of
uncertainty. For a more detailed treatment see \citet{2009MNRAS.395.2127O}.

The mean number density of objects with mass $m_{\BH}$ in a dynamically relaxed galactic nucleus
around a supermassive black hole (SMBH) of mass $M_{\SMBH}$
is
\begin{equation}\label{e:n}
\langle n_{\BH}(r)\rangle = \frac{3-\alpha_{\BH}}{4\pi} \frac{2 \kappa_{\BH} M_{\SMBH}}{m_{\BH}} r_i^{-3} \left(\frac{r}{r_i}\right)^{-\alpha_{\BH}}\,,
\end{equation}
where the total mass of stars within the radius of influence, $r_i$, is $2M_{\SMBH}$,
the total mass fraction in $m_{\BH}$ is $\kappa_\BH$, and
\begin{equation}\label{e:ri}
r_i = \frac{G M_{\SMBH}}{\sigma^2} = \frac{G M_0}{\sigma_0^2} \left( \frac{M_{\SMBH}}{M_0}\right)^{1-(2/k)}\,,
\end{equation}
where we have used $M_{\SMBH} = M_0 (\sigma/\sigma_0)^{k}$ according to the $M_{\SMBH}-\sigma$ relation, in which
$\sigma_0 = 200\, \rm km/s$, $M_0=1.3\times 10^8\Msun$, and  $k=4$ \cite{2008gady.book.....B,2009ApJ...698..198G}.
We extrapolate this density profile inwards until a radius
which encloses only 1 BH,
\begin{equation}\label{e:rmin}
 r_{\BH\min}=r_{\BH1} = N_{\BH}^{-1/(3-\alpha_{\BH})} r_i\,,
\end{equation}
where $N_{\BH} = 2 \kappa_\BH M_{\SMBH}/m_{\BH}$ is the number of BHs within $r_i$.

The maximum impact parameter, $b_{\max}$ for a GW capture for given relative velocity $w$ is from Eq.~(17) of \citet{2009MNRAS.395.2127O}
\begin{equation}\label{e:bmax}
 b_{\max} = \frac{G}{c^2}\left( \frac{340\pi}{3} \right)^{1/7} \frac{\eta^{1/7}}{w^{9/7}} m_{\BH \rm tot}\,,
\end{equation}
where $\eta$ is the symmetric mass ratio and $m_{\BH \rm tot}$ is the total binary mass. We will take equal mass binaries for which
$\eta=1/4$ and $m_{\BH \rm tot}= 2m_{\BH}$.
In the following we assume for simplicity that the relative velocity is the circular velocity at radius $r$ from the SMBH,
 $w=v(r)=(M_{\SMBH}/r)^{1/2}$.

The scattering cross section is $\sigma_{\rm cs} = \pi b_{\max}^2$. The binary capture rate in a spherical shell of thickness $\D r/r$ is
\begin{align}
 \frac{\D}{\D\ln r}\Gamma &=  (4\pi r^3) n_{\BH}^2 \sigma_{\rm cs} v  \nonumber\\
&=  {\Gamma}_0 \kappa_{\BH}^2
\left(\frac{M_{\SMBH}}{M_0}\right)^{31/(7 k) - 1}
\left(\frac{r}{r_i}\right)^{(53/14) - 2\alpha_{\BH}}\,,
\label{e:dGamma}
\end{align}
in the second equation we have plugged in Eqs.~(\ref{e:ri}--\ref{e:bmax}),  and
\begin{align}\label{e:dGamma0}
 \Gamma_0 &= 4 \left(\frac{340\pi \eta}{3}\right)^{2/7} (3-\alpha_{\BH})^2 \left(\frac{\sigma_0}{c}\right)^{31/7} \frac{c^3}{G M_0}\xi\\\nonumber
&=6.13 \times 10^{-9}\yr^{-1}\times(3-\alpha_{\BH})^2 \xi\,.
\end{align}
where $\xi=n_{\BH}^2(r_i) /\langle n_{\BH}(r_i)\rangle^2$.
From Eq.~(\ref{e:dGamma}) it is clear that for fixed $r/r_i$, the rates are independent of $m_{\BH}$, if their total mass fraction
in the cluster, $\kappa_\BH$, is fixed. This follows
from the fact that  $\D\Gamma/\D\ln r\propto n_{\BH}^2 b_{\max}^2$  where
$\langle n_{\BH}\rangle^2 \propto m_{\BH}^{-2}$ and $b_{\max}^2 \propto m_{\BH}^2$, see Eqs.~(\ref{e:n}) and (\ref{e:bmax}).

Integrate $\D\Gamma/\D r$ between $r_{\BH\min}\ll r_i$ to get the total rate in one galaxy, using Eq.~(\ref{e:rmin}),
\begin{align}
 \Gamma_{1\rm GN} &=  \Gamma'_0 \kappa_{\BH}^2
\left(\frac{M_{\SMBH}}{M_0}\right)^{31/(7 k) - 1} \left[ \left(\frac{r_{\BH\min}}{r_i}\right)^{(53/14) - 2\alpha_{\BH}} -1 \right]\nonumber\\
&\approx \label{e:Gamma1}
 \Gamma''_0 \,
 \kappa_{\BH}^\frac{31}{14(3-\alpha_{\BH})} \left(\frac{M_{\SMBH}}{M_0}\right)^{\frac{31}{7 k} - 1}
\left(\frac{2M_{\SMBH}}{m_{\BH}}\right)^{ \frac{2\alpha_{\BH} - (53/14)}{3-\alpha_{\BH}}}
\end{align}
where $\Gamma'_0 = [ 2\alpha_{\BH}-53/14]^{-1} \Gamma_0 $ and
\begin{equation}\label{e:Gamma2}
 \Gamma''_0 = 4 \left(\frac{340\pi \eta}{3}\right)^{2/7} \frac{(3-\alpha_{\BH})^2}{2\alpha_{\BH} - (53/14)}
\left(\frac{\sigma_0}{c}\right)^{31/7} \frac{c^3}{G M_0} \xi\,.
\end{equation}
Let us introduce normalized parameters with numbers representative of a Milky-Way-size galaxy,
$\bar{\kappa}_{\BH}=\kappa/0.025$, $M_{\SMBH} = M_{4e6} \times 4 \times 10^6\Msun$, $m_{\BH} = m_{\BH1}\times10\Msun$,
$r_{\BH\min}= r_i/ N_\BH$.
\begin{align}\label{e:Gamma}
 \Gamma_{1\rm GN}
&\approx
3.0\times 10^{-9}\yr^{-1}\,
 \xi_{30}\, \bar{\kappa}_{\BH}^{31/14} \,m_{\BH 1}^{ -3/14} \,M_{4e6}^{9/28}
\end{align}
where $\xi = 30\, \xi_{30}$, and we assumed $\alpha_{\BH}=2$.
The rate per Milky-Way galaxy ($\bar{\kappa}_{\BH}=M_{4e6}=m_{\BH 1}=1$) is
around $3\times 10^{-9}\xi_{30} \yr^{-1}$.

The total rate is the sum of the rates of individual galaxies within a detectable distance, $d_{\max}$,
\begin{align}\label{e:rate}
 \Gamma &= \frac{4\pi }{3}d_{\max}^{3} n_{\rm gal} \Gamma_{1\rm GN} \nonumber\\
&=
4.5 \yr^{-1}\,
 \xi_{30}\, \bar{\kappa}_{\BH}^{31/14}\,m_{\BH 1}^{ -3/14} \,M_{4e6}^{9/28} n_{\rm gal, 5} d_{\max, 2}^{3}
\end{align}
where $d_{\max 2} = d_{\max,  2}\times 2\Gpc$, and $n_{\rm gal} = n_{\rm gal, 5}\times
0.05\Mpc^{-3}$.  This is comparable to the rates found in Table 1 of
 \citet{2009MNRAS.395.2127O}.\footnote{ To directly compare the rates in Table~1 of \citet{2009MNRAS.395.2127O},
the values should be scaled up by a factor of $\sim 2$ to
account for the slightly smaller normalization used in that paper.}

The take-away from these calculations can be summarized as follows.
\begin{itemize}
 \item In  \citet{2009MNRAS.395.2127O}, Fig. 11, shows that the maximum detection limit of the full signal
(including the RB phase and the final chirp)
is between $1$--$3\,$Gpc for $10\Msun\lesssim m_{\BH}\lesssim 500\Msun$ for a broad range of impact parameters
for Advanced LIGO with $S/N=5$.
Eq.~(\ref{e:Gamma}) shows that the rates are weakly sensitive to $m_{\BH}$ as long as their total mass in the cluster is fixed.
The event rates involving a few IMBHs may be
equally numereous as rates among many stellar mass BHs with the same total mass.
 \item \citet{2009MNRAS.395.2127O} have shown that the event rates are dominated by close first encounters, where
the signal enters the LIGO band in the RB phase. This is to be expected as the event rates
are dominated by the innermost objects, where the velocity dispersion is large, requiring a close approach for
binary formation.
 \item Eq.~(\ref{e:Gamma}) shows that the rates are weakly sensitive to $M_{\SMBH}$.
This is in stark contrast to other gravitational wave sources
whose rates scale proportionally to the mass of the galaxy.
The net rates are determined by the number density of lower mass galaxies
which greatly outnumber Milky-Way sized galaxies.
 \item Observations show that many galaxies can significantly deviate from the $M-\sigma$ relation, in terms of $n(r_i)$.
This implies that the average value of
$\langle \xi\rangle \equiv \langle n_{\BH}^2\rangle / \langle n_{\BH}\rangle ^2 \gg 1$ in Eqs.~(\ref{e:dGamma0}) and (\ref{e:Gamma2}).
If the RMS scatter in the densities from the $M-\sigma$ relation is $5$, then $\xi = 25$.
 \item The exact average value of $\kappa_{\BH}$ and $\alpha_{\BH}$ is
   uncertain, due to uncertainties in the final initial mass function
   in these environments. While $\kappa_{\BH}=2.5\%$ ($\bar\kappa=1$)
   may be reasonable for the center of the Milky Way, assuming $20,000$
   BHs of mass $10\Msun$ (see refereces in \cite{2009MNRAS.395.2127O}), other values may also be possible in
   general. Indeed, the relaxation timescale in M32, which is a dwarf elliptical galaxy hosting a
   $2\times 10^6\,\Msun$, is short enough that black holes can segregate from
   a larger volume of stars than in the Milky Way. In  \citet{2009MNRAS.395.2127O},
   we direclty solved for $\kappa_{\BH}$ and $\alpha_{\BH}$,
   using the average density of stars at the radius of influence and
   the chosen initial mass function of the stars. Recent observation
   of the Galactic nucleus shows evidence for an extremely top-heavy
   mass function \cite{2010ApJ...708..834B}.  Equation~(\ref{e:rate}) shows that the rates are
   strongly dominated by the fraction of galaxies with relatively
   large values of $\kappa_{\BH}$. For example, galactic nuclei with a
   $\kappa_{\BH}=50\%$ mass--fraction of compact objects (so that
   $\bar{\kappa}=20$) contribute $\bar{\kappa}_{\BH}^{31/14} = 760$
   times larger event rate than the nominal Milky Way estimate with $\bar{\kappa}=1$.  This explains the larger values of event
   rates in models E and F in O'Leary et al. (2009).
 \item Other effects may further increase the event rates, which were not included in O'Leary et al. (2009).
Keshet, Hopman, \& Alexander (2009) found that galactic nuclei dominated by light objects can lead to a steeper
density profile for the massive objects with $2<\alpha_{\BH}<3$. Eq.~(\ref{e:Gamma1}) shows that the rates are exponentially sensitive
to $\alpha_{\BH}$, these larger values lead to much higher rates for the larger mass objects.
Further, mass segregation in vector resonant relaxation leads to an anisotropic configuration of compact objects, which
increases the rates by the square of the linear flattening of the distribution.
\end{itemize}

}

\bibliography{ms}
\end{document}